\documentclass[fleqn,usenatbib]{mnras}

\usepackage[T1]{fontenc}

\DeclareRobustCommand{\VAN}[3]{#2}
\let\VANthebibliography\thebibliography
\def\thebibliography{\DeclareRobustCommand{\VAN}[3]{##3}\VANthebibliography}

\usepackage{graphicx}	
\usepackage{amsmath}	
\usepackage{amssymb}	
\usepackage[normalem]{ulem}
\usepackage{tablefootnote}
\usepackage{pdflscape}
\usepackage{lscape}
\usepackage{gensymb}

\title[Source region of waves observed in coronal fan loops]{Exploring source region of 3-min slow magnetoacoustic waves observed in coronal fan loops rooted in sunspot umbra}

\author[Rawat \& Gupta]{Ananya Rawat$^{1,2}$\thanks{E-mail: ananyarawat@prl.res.in (AR)}, Girjesh R. Gupta$^{1}$\thanks{E-mail: girjesh@prl.res.in (GRG)}
\\
$^{1}$ Udaipur Solar Observatory, Physical Research Laboratory, Dewali, Badi Road, Udaipur 313001, India \\
$^{2}$ Department of Physics, Indian Institute of Technology Gandhinagar, Palaj, Gandhinagar 382355, India \\
}

\date{Accepted . Received ; in original form}

\pubyear{2023}

\begin{document}
\label{firstpage}
\pagerange{\pageref{firstpage}--\pageref{lastpage}}
\maketitle

\begin{abstract}

Sunspots host various oscillations and wave phenomena like umbral flashes, umbral oscillations, running penumbral waves, and coronal waves. All fan loops rooted in sunspot umbra constantly show a 3-min period propagating slow magnetoacoustic waves in the corona. However, their origin in the lower atmosphere is still unclear. In this work, we studied these oscillations in detail along a clean fan loop system rooted in active region AR12553 for a duration of 4-hour on June 16, 2016 observed by Interface Region Imaging Spectrograph (IRIS) and Solar Dynamics Observatory (SDO). We traced foot-points of several fan loops by identifying their locations at different atmospheric heights from the corona to the photosphere. We found presence of 3-min oscillations at foot-points of all the loops and at all atmospheric heights. We further traced origin of these waves by utilising their amplitude modulation characteristics while propagating in the solar atmosphere. We found several amplitude modulation periods in the range of 9--14 min, 20--24 min, and 30--40 min of these 3-min waves at all heights. Based on our findings, we interpret that 3-min slow magnetoacoustic waves propagating in coronal fan loops are driven by 3-min oscillations observed at the photospheric foot-points of these fan loops in the umbral region. We also explored any connection between 3-min and 5-min oscillations observed at the photospheric foot-points of these loops and found them to be weakly coupled. Results provide clear evidence of magnetic coupling of the solar atmosphere through propagation of 3-min waves along fan loops at different atmospheric heights.

\end{abstract}



\section{Introduction}
\label{sec:intro}

Magnetohydrodynamic (MHD) waves  observed in the solar atmosphere are studied mainly for their role in coronal heating and atmospheric seismology \citep{2012RSPTA.370.3193D}. At different atmospheric heights, sunspots show different features that host various oscillations and waves. There are numerous observational reports of waves and oscillations above the sunspot umbra in the photosphere, umbral ﬂashes and running waves in the chromosphere, and propagating waves in the corona. However, several open questions still exist related to studies of waves and oscillations in the sunspots which need to be addressed and clarified  \citep[for more details see, e.g.][]{2015LRSP...12....6K,2016PhDT........15L}.

At the photosphere, sunspots usually show strong power in  5-min oscillations and significant power in 3-min oscillations bands \citep{2000ApJ...534..989B,2021RSPTA.37900175N}. In the chromospheric sunspot umbra, sudden strong brightenings occur at random locations with a period of about 3-min, which are called umbral flashes \citep{1969SoPh....7..351B}. These flashes are strongly non-linear and have asymmetric light curves (saw-tooth shape), and are interpreted as signatures of upward propagating magnetoacoustic shock waves \citep[e.g.][]{2006ApJ...640.1153C}. Such shock wave behaviours are also observed in the transition region \citep{2014ApJ...786..137T}. Usually, at the umbral photosphere, 5-min oscillations dominate over 3-min oscillations, but at heights above the temperature minimum, 3-min oscillations dominate due to acoustic cut-off of 5-min oscillations \citep[e.g.][]{2016PhDT........15L}. \citet{2010ApJ...719..357F} were able to reproduce such period shifts with height using 3-D numerical simulations. Various attempts have been made to connect the photospheric oscillations with umbral flashes and chromospheric running waves.  \citet{2015ApJ...800..129M} suggested that both umbral flashes and running waves originate from photospheric p-mode (5-min) oscillations. However, \citet{2017ApJ...836...18C} suggested that 3-min chromospheric oscillations in the sunspot are generated by photospheric 3-min oscillations produced by magnetoconvection happening inside the umbral dots and light bridges.

 Loop-like coronal structures rooted in the sunspot umbra show outward propagating disturbances with subsonic phase speed and period around 3-min \citep[e.g.][]{2002A&A...387L..13D,2020A&A...638A...6S}. Similar propagating disturbances with period around 15 min are also observed along plume like structures in the polar coronal holes \citep[e.g.][]{1998ApJ...501L.217D,2010ApJ...718...11G}. These propagating disturbances are found to have wave-like properties, and are often interpreted in terms of propagating slow magnetoacoustic waves \citep[e.g.][]{2012SoPh..279..427K,2012A&A...546A..93G}. 3-min slow magnetoacoustic waves propagating along coronal fan loops show amplitude modulations in the period range of 20-30 min \citep{2020A&A...638A...6S}. Such amplitude modulations are result of interaction of various beat like phenomena formed due to the number of closely spaced frequencies within the 3-min period band \citep[e.g.,][]{2006ApJ...643..540M,2015ApJ...812L..15K,2020A&A...638A...6S}. Formation of such closely spaced multiple frequencies within the 3-min period band are explained either as eigenvalues of umbral oscillations or through the resonant filtering mechanism by \citet{2005A&A...433.1127Z}. Although multiple frequencies exist within the 3-min period band, reports of modulation period is limited to only isolated periods within 20-30 min range \citep[e.g.,][]{2015ApJ...812L..15K,2020A&A...638A...6S}. 

  Although propagating coronal slow waves are ubiquitous in the different structures, observational evidence of their source region is still rare \citep[e.g.][]{2012ApJ...757..160J,2015ApJ...812L..15K}.   \citet{2012ApJ...757..160J} found 3-min magnetoacoustic waves in the coronal fan loops, which were rooted in the umbral dots at the photosphere.  In these umbral dots, the power of 3-min oscillations were enhanced compared to surrounding regions.  \citet{2015ApJ...812L..15K} utilised the amplitude modulation of 3-min Fourier-filtered light curves obtained at near the foot-point of the fan loop at different atmospheric layers above the sunspot umbra, and associated the presence of 3-min slow magnetoacoustic waves in corona with the 5-min photospheric p-mode. Similarly, \citet{2016ApJ...830L..17Z} tracked 5-min p-mode waves from the photosphere to the  corona in active regions using a time-distance helioseismology analysis technique. \citet{2017ApJ...850..206S} reported the influence of umbral flashes on different sunspot waves in the upper atmosphere based on their synchronized change in the amplitude of 3-min oscillations. On the other hand, \citet{2013A&A...554A.146K} found no connection between 3-min oscillations and coronal fan structures in AIA 171 \AA\ passband. Henceforth, there are mixed opinions on the origin of these 3-min waves in the umbra based on a few reports. Moreover, contrary views exist on 3-min and 5-min oscillations observed at the photosphere. \citet{1986ApJ...301..992L} concluded that 5-min oscillations do not drive 3-min oscillation in the umbra because 3-min chromospheric oscillations are not correlated with the 5-min photospheric oscillations. \citet{2015ApJ...812L..15K} suggested that 5-min oscillations are responsible for the 3-min waves present in the coronal loops whereas \citet{2017ApJ...836...18C} suggested that these are not coupled in the umbra due to their different origins. Therefore, to obtain a more general understanding of how these 5-min photospheric p-mode oscillations compare with 3-min waves, a detailed investigation is needed.

For direct and unambiguous detection of waves, it is mandatory to have excellent signals at different atmospheric layers, which is not always the case. Here, we present a multi-wavelength analysis of propagation of slow magnetoacoustic waves from the photosphere to the corona by studying fan loop structures anchored within a sunspot umbra. Fan loop structures provide an excellent site to study the propagation of 3-min slow  magnetoacoustic waves in the solar atmosphere. Our motivation is to investigate whether the 3-min waves present in the corona owe their origin to the photosphere or not. Therefore, this study will also probe the magnetic connectivity of the whole solar atmosphere. We present the details of observation in Section~\ref{sec:obs}, data analysis and results in Section~\ref{sec:analysis}, and finally discuss and summarise our results in Section~\ref{sec:discussion}.

\section{Observations}
\label{sec:obs}

\begin{figure*}
    \centering
    \includegraphics[width=0.9\textwidth]{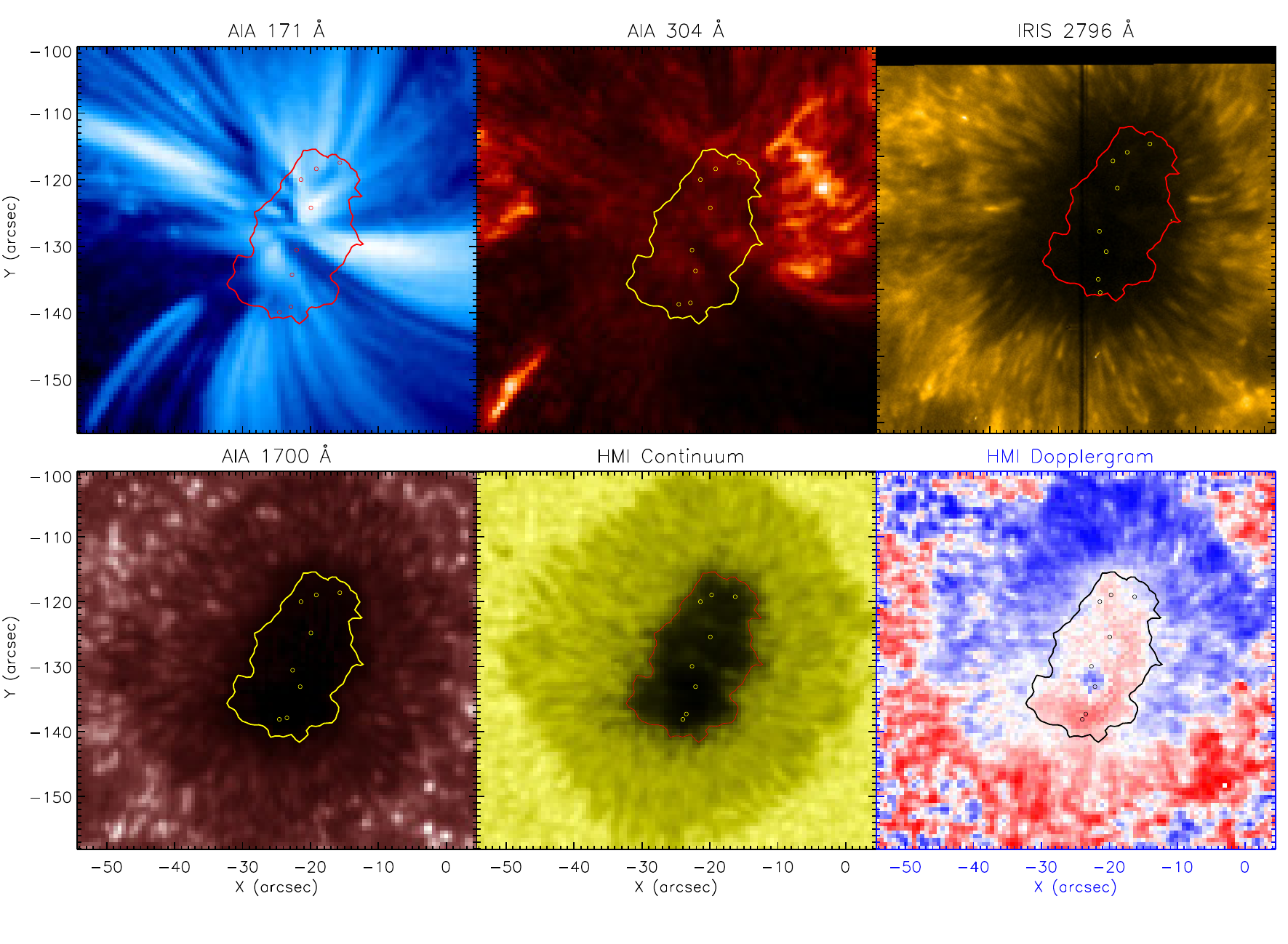}
    \caption{Images of sunspot and fan loops as obtained from different AIA, IRIS, and HMI passbands as labelled. Small circles (o) over different panels represent location of all the identified loops at that atmospheric height (details in Figs.~\ref{fig:map171} and \ref{fig:alignment}), which are used for detailed analysis. Contours over different panels indicate the umbra-penumbra boundary as obtained from the HMI continuum image.}
    \label{fig:maps}
\end{figure*}

To investigate the origin and characteristics of waves, we are using multi-wavelength observations of cool fan loops observed within the active region. For this purpose, we have identified an appropriate data set observed by Atmospheric Imaging Assembly \citep[AIA;][]{2012SoPh..275...17L}, Helioseismic and Magnetic Imager \citep[HMI;][]{2012SoPh..275..207S} both onboard Solar Dynamics Observatory \citep[SDO;][]{2012SoPh..275....3P}, and Interface Region Imaging Spectrograph \citep[IRIS;][] {2014SoPh..289.2733D}. To study waves at the photosphere, we utilise intensity continuum and Dopplergram images obtained from HMI. AIA EUV images provide good coverage over the transition region and the corona. To obtain good coverage over the chromosphere, we are utilising UV images obtained from AIA and IRIS. Sunspot studied here belongs to the active region NOAA AR 12553 (7$^{\circ}$S, 8$^{\circ}$W) which was observed on June 16, 2016. We obtained 4-hours of data starting from 07:19:11 UT, as shown in Fig.~\ref{fig:maps}. 

AIA/SDO provides full-disk solar images in seven EUV channels manifesting the upper atmosphere and three UV-visible channels manifesting the lower atmosphere. HMI/SDO provides full disk images of the photospheric Sun in intensity continuum, Dopplergram, and magnetogram which are derived using Fe I 6173 \AA\ spectral line. These instruments are recording continuous images of the Sun since its launch. All the images were calibrated, co$-$aligned, and re-scaled to a common 0.6$\arcsec$/pixel resolution, and 12 s temporal resolution using the robust SDO library of Rob Rutten\footnote{https://robrutten.nl/rridl/00-README/sdo-manual.html}. This tool incorporates standard $aia\_prep.pro$ and $hmi\_prep.pro$ routines available in the standard Solarsoft (SSW), and also aligns images from multiple filters and corrects for any time-dependent shifts.

Active region was also observed by IRIS in 2-step raster mode for more than 4-hours starting from 07:19:13 UT. IRIS provided Slit-Jaw-Images (SJI) in 2796 \AA\ and 1400 \AA\ passbands. IRIS-SJI 2796 \AA\ passband manifests a chromospheric temperature of 10,000 K due to coverage of Mg II  line. Due to the poor signal in IRIS 1400 \AA\ images, we are not utilising it in our analysis. Images obtained from IRIS-SJI 2796 \AA\ passband have an exposure time of 2 s with an effective cadence of 6.88 s, 0.166$\arcsec$/pixel resolution, and field-of-view of $60\arcsec\times 65\arcsec$\ centered around the sunspot umbra. In the later part of observation, several data gaps were found. Therefore, we selected only the first 4 hours of continuous data, without any data-gap to perform our analysis. 

Imaging data allows us to co-align data from different instruments, thus allowing us to simultaneously use data from different instruments working at different wavelengths. We co-aligned IRIS and SDO observations using IRIS-SJI 2796 \AA\  and AIA 1700 \AA\ images using cross-correlation method. All IRIS, AIA and HMI images are derotated with respect to time at 07:19:17 UT using the SSW routines. The identified data set provides a unique opportunity to study spatial and temporal evolution of waves in the plane of sky along the whole solar atmosphere.

Since observed sunspot is slightly off the disk-center (heliocentric co-ordinates $X\approx -25\arcsec, Y\approx -125\arcsec$), the angle between vertical and LOS is $\approx 7.7\degree$ which leads to $ \mu=cos \theta \approx 0.99$. So, any projection effect on Dopplergram velocity oscillations and other parameters will be almost negligible. In the umbral region, plasma-$\beta=1$ layer is below the photosphere ($\tau_{5000}=1$) and at this layer fast waves get converted into slow waves \citep{2015ApJ...807...20P}, and thus enables us to detect slow waves from the HMI continuum data. We studied the sunspot in AIA 193 \AA, AIA 171 \AA, AIA 304 \AA, IRIS 2796 \AA, AIA 1700 \AA, and HMI continuum and Dopplergram images manifesting coronal temperature of 1.6 MK, 0.7 MK, transition region temperature of 50,000 K, chromospheric temperature of 10,000 K, temperature minimum region of 5000 K, and photospheric temperature of 6000 K, respectively, covering the different layers of solar atmosphere. Fig.~\ref{fig:maps} shows the image of sunspot observed from different passbands as labelled. It is clear from the images that sunspot looks quite different at different atmospheric layers which highlights involved complexities in their dynamics.

\section{Data Analysis and Results}
\label{sec:analysis}

 \begin{figure}
    \centering
    \includegraphics[width=0.5\textwidth]{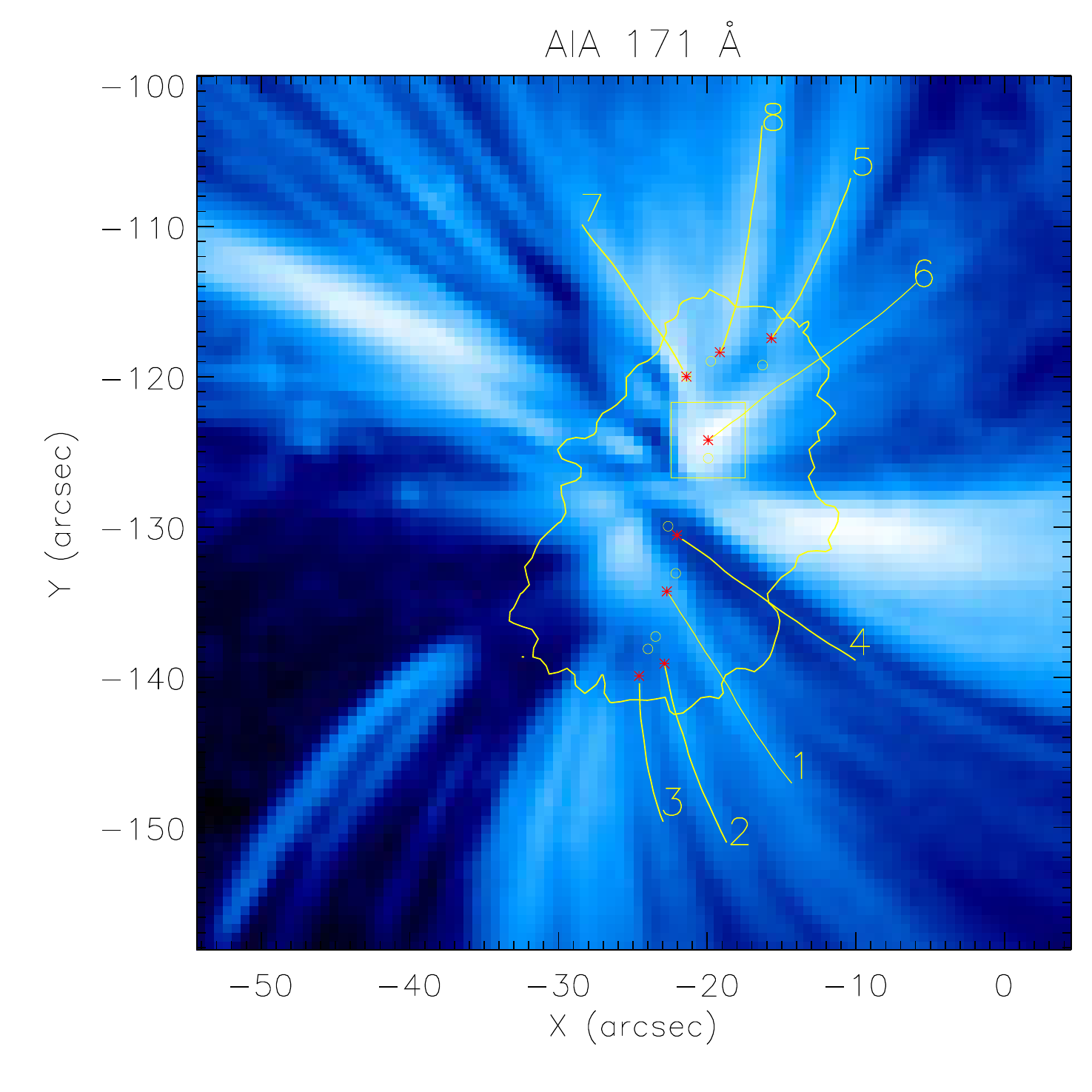}
    \caption{Fan loop system observed in AIA 171 \AA\ passband. Identified loop foot-points in the corona and photosphere are marked with asterisks (*) and circles (o) respectively, which are used for detail analysis. Traced coronal loops are also drawn for visualisation purpose only. Contour indicates umbra-penumbra boundary as obtained from HMI continuum image. Box enclosing the coronal foot-point of loop 6 indicates the region chosen for image correlation analysis to identify locations of the loop in the lower atmosphere.}
    \label{fig:map171}
\end{figure}

Fig.~\ref{fig:map171} shows the analysed fan loop structures in AIA 171 \AA\ passband. Overplotted contour represents the umbral boundary of the sunspot obtained from the HMI continuum image at the count of 9000 DN. Asterisks (*) represent the coronal foot-point of fan loops in the AIA 171 \AA\ passband. This location will be used as a reference while identifying the location of these loops in the lower solar atmosphere where loops are not distinguishable. We have identified eight fan loops emanating from the sunspot umbra for our study purpose and labelled them accordingly. Associated coronal loops are also drawn for visualisation purpose only. Properties of slow magnetoacoustic waves propagating along some of these coronal fan loops are described and discussed in \citet{2020A&A...638A...6S}. In this work, we analyse all the identified loop locations at different atmospheric heights. However, here we present results from loop 6 as a representative example. Results from all the other loops are summarized in Section~\ref{section:all}. 
 
\subsection{Identification of loop locations in the lower atmosphere}
\label{sec:correlation}

 \begin{figure}
    \includegraphics[width=0.49\textwidth]{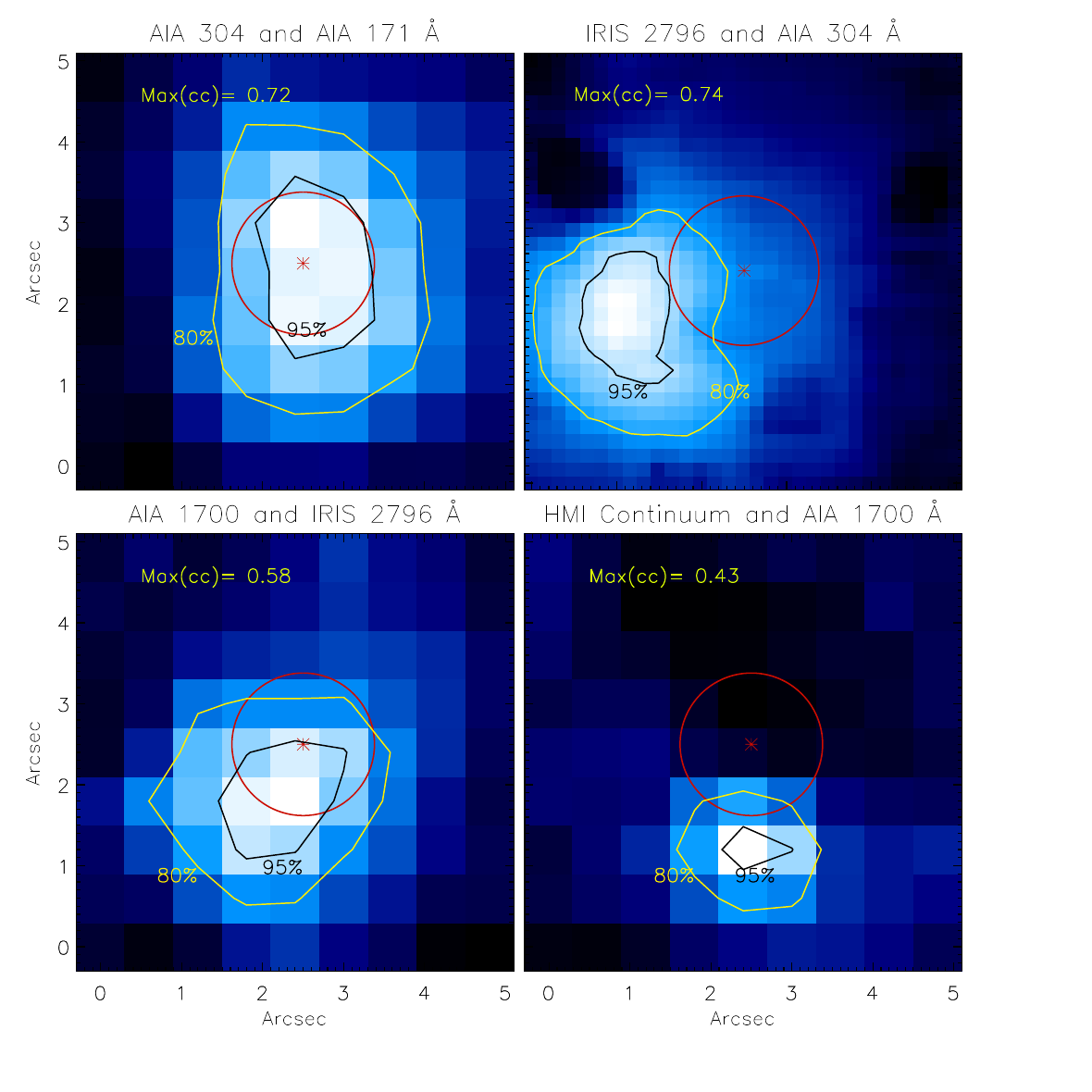}
    \caption{Correlation images obtained between various atmospheric heights as labelled. In each panel, the asterisk sign (*) in the center refers to the coronal foot-point of loop 6. The red circle in all the panels represents the cross-section of the loop obtained from AIA 171 \AA\ image (see details in appendix~\ref{append:area}). Overplotted black and yellow color contours are obtained at $\approx$ 95\% and 80\% of maximum correlation values respectively.} 
    \label{fig:alignment}
\end{figure}

\begin{figure*}
    \centering
    \includegraphics[width=0.9\textwidth]{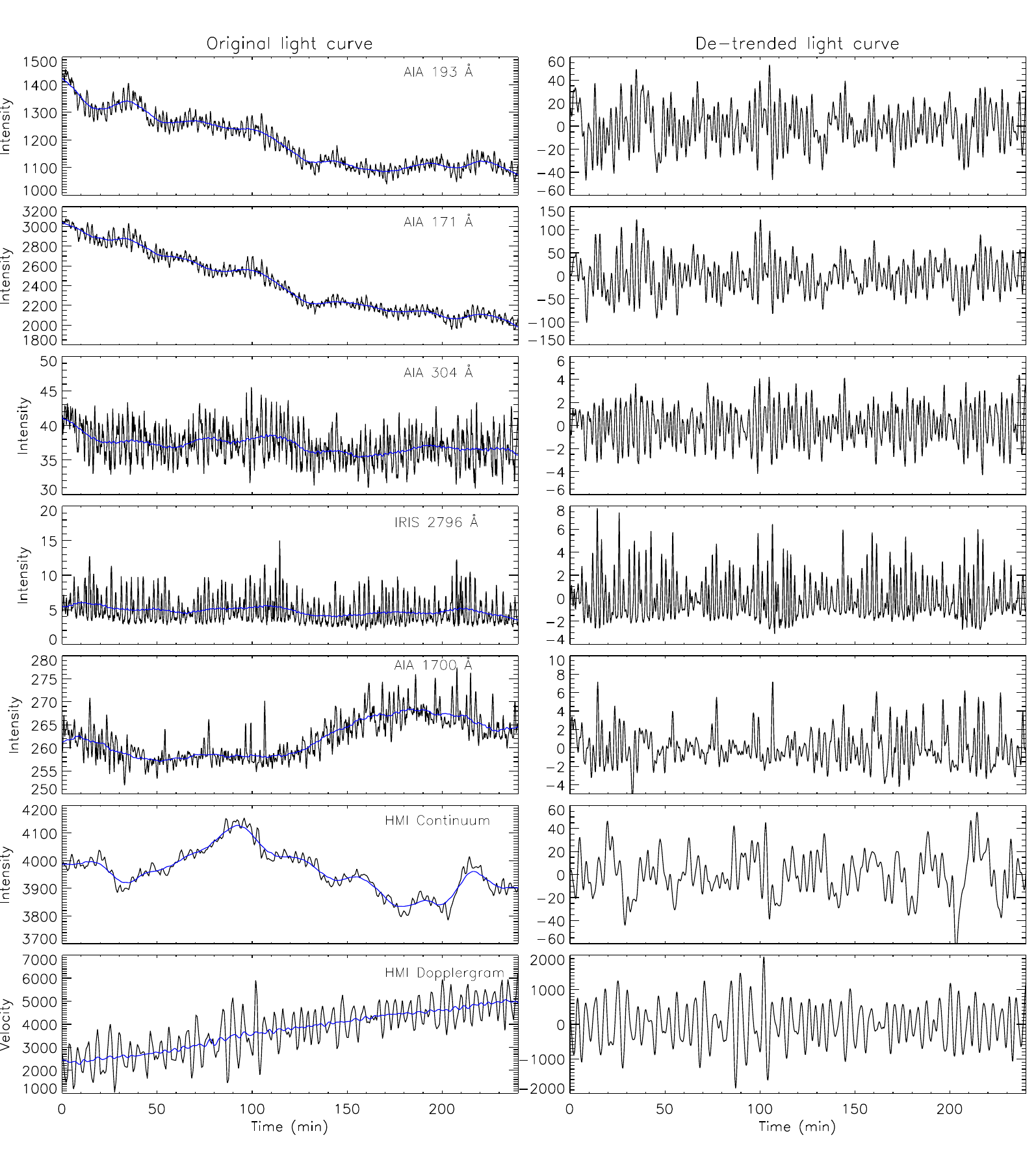}
    \caption{The left and right panels show original and background subtracted smoothened light curves for loop location 6 in AIA 193 \AA, AIA 171 \AA, AIA 304 \AA, IRIS 2796 \AA, AIA 1700 \AA, HMI continuum and HMI Dopplergram as labelled. The overplotted blue lines in the left panels show the background trend at each layer of the solar atmosphere of loop location 6.}
    \label{fig:lightcurve}
\end{figure*}

\begin{figure*}
    \centering
    \includegraphics[width=0.9\textwidth]{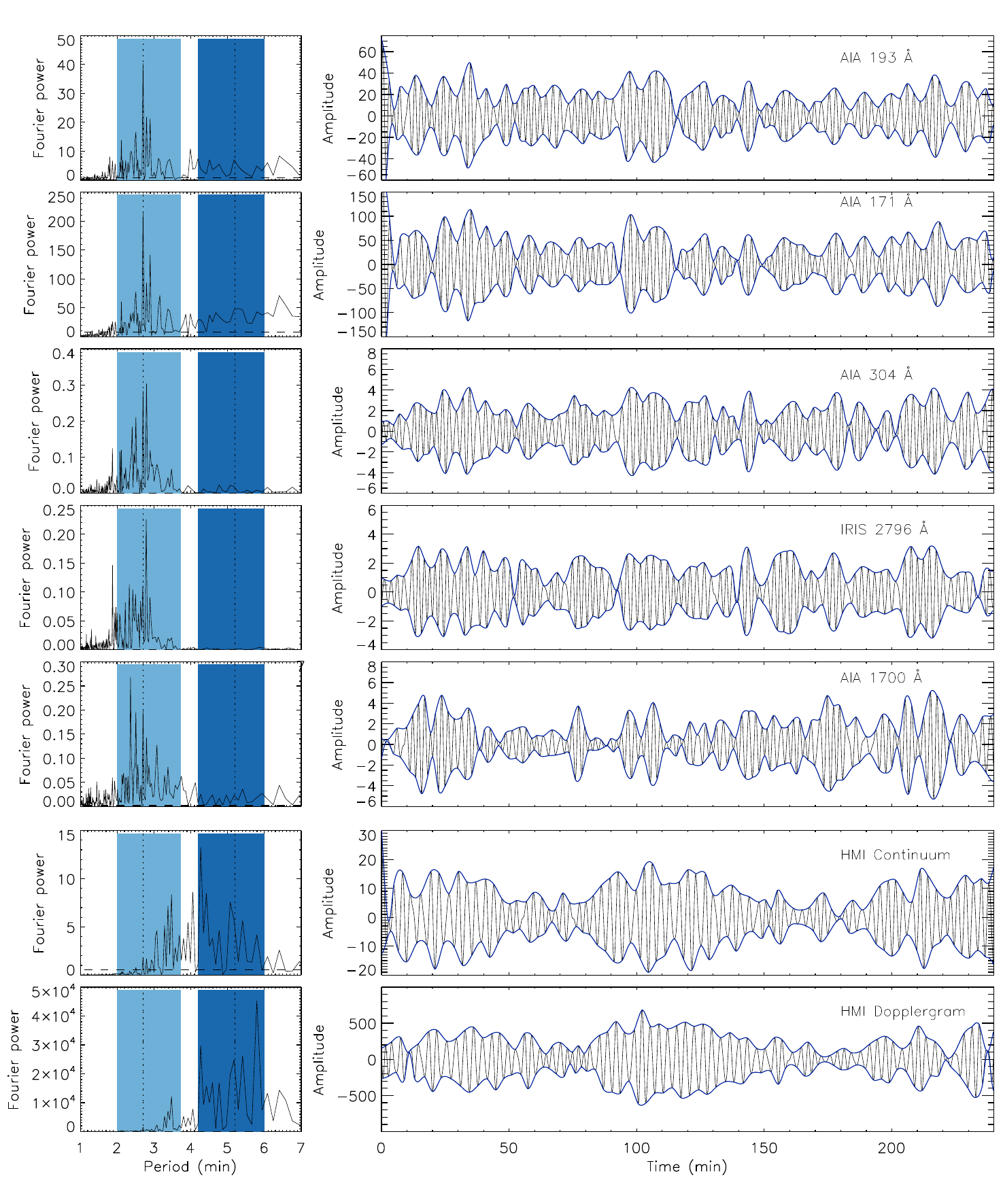}
    \caption{Left panels: FFT power spectrum of original light curves obtained at each location of loop 6 in the solar atmosphere as shown in the left panels of Fig.~\ref{fig:lightcurve}. Shaded regions in light blue colour denote the 3-min period band used to produce the filtered light curves shown in the right panels. Shaded regions in dark blue colour denote the 5-min period band. Horizontal dashed lines represent 95\% significance level. Vertical dashed lines represent the dominant peak in the 3-min and 5-min period band present at all the atmospheric heights and at the photosphere respectively. Right panels: 3-min filtered light curves with over-plotted blue lines show amplitude modulation envelopes traced using maxima and minima of respective filtered light curves.}
    \label{fig:filtered}
\end{figure*}

To determine the cross-sectional area of coronal fan-loop in AIA 171 \AA , we obtained intensity across the coronal foot-point marked by asterisk sign (*) in Fig.~\ref{fig:map171} and its neighbouring pixels, and fitted it with a Gaussian profile. We then extracted the Full-Width Half Maxima (FWHM) of this Gaussian and considered it as the diameter of the loop $\approx 1.76\pm 0.18\arcsec$ ($1.27\pm0.13$ Mm)  as described in the Appendix~\ref{append:area}.

Similarly, for lower heights where the loop is not visible, we perform correlation analysis to determine the loop location, and also if possible then its cross-sectional area.  Here, we choose a box of size $5\arcsec \times 5\arcsec$  by keeping the location of loop 6 in AIA 171 \AA\ image in center as indicated by an asterisk sign (*) in the Fig.~\ref{fig:alignment}. To perform the cross-correlation, we utilised 16-min background subtracted and 1-min smoothed light curves so as to remove any low-frequency background and high-frequency noise from the light curves. This makes correlations to be only depend on 3-min oscillations and its related variations. We have interpolated the AIA 304 and 1700 \AA\ light curves at the cadence of 6.88 s to match with the cadence of IRIS SJI 2796 \AA\ using IDL routine $interpol$ for the purpose of correlation. For the correlation between AIA 1700 \AA\ and HMI continuum light curves, we are utilising 3-min filtered light curves for both passbands to avoid any influence of 5-min oscillations observed in the HMI continuum. We noted the maximum correlation coefficient at each pixel, and created a cross-correlation image as shown in Fig.~\ref{fig:alignment}. Therefore, the image provides maximum correlation coefficient values at each pixel. It should also be noted that the image appears similar if we utilise 3-min filtered light curves for all the passbands. We obtain a black colour contour over this correlation image at 95\% of maximum correlation value within the image whereas the red circle represents the cross-section of the loop in the corona as obtained from AIA 171 \AA\ image. Choice of contour level at 95\% of maximum correlation value is obtained by comparing correlation images in the corona with FWHM of loop cross-section as described in the Appendix~\ref{append:corr171}. 

 Central pixel of black contour in the correlation images of Fig.~\ref{fig:alignment} depicts location of the loop at that atmospheric height. This location has been further utilised to identify loop locations at the lower heights from the cross correlation of light curves as described above. In this way, we are able to trace the same source of 3-min oscillation from the corona to the photosphere. Since these 3-min waves can propagate from photosphere to corona only through the wave-guide, black contours here may represent either a single loop or collection of loops emanating from the same photospheric oscillating region. Although chosen contour level may possibly depict the loop cross-section at that height, this needs a detailed investigation which is beyond the scope of current work and is a topic of future interest. Henceforth, we assume that the obtained 95\% correlation contour region signifies coherent oscillating region of 3-min oscillations which may provide useful constraints on the upper limit on the size of loop cross section at that height. Size of the loop or oscillating region mainly depends on choice of contour level. Therefore, we also obtained contour levels at 80\% of maximum correlation value. These contours cover the correlation patches observed in correlation images very nicely, and result in quite broader oscillating regions. However, more importantly, for any choice of contour levels, we find that the size of the coherent oscillating region decreases as we move into the lower atmosphere.  This is expected as per the theory of flux tube expansion with height \citep[e.g.,][]{2004psci.book.....A}. However as mentioned earlier, analysis demands a detailed investigation to provide an estimate on the size of loops.
 
 Nevertheless, we obtained area of 95\% and 80\% contour levels in  AIA 304 \AA , IRIS 2796 \AA , AIA 1700 \AA , and HMI continuum passbands by visually fitting the contours with either circle or ellipse depending on their shape. Obtained areas are $\approx$ $2.03\pm1.16$ ($1.07\pm0.61$), $1.24\pm0.23$ ($0.65\pm0.12$), $1.16\pm0.74$ ($0.61\pm0.45$), and $0.18\pm0.32$ ($0.09\pm0.16$) arcsec$^2$  (Mm$^2$), respectively for 95\% contour level whereas that for 80\% contour levels are $\approx$ $7.46\pm2.09$ ($3.92\pm1.1$), $6.78\pm0.61$ ($3.56\pm0.32$), $5.45\pm1.74$ ($2.86\pm0.91$), and $2.06\pm1.07$ ($1.08\pm0.56$) arcsec$^2$ (Mm$^2$) respectively. Error bars are calculated by assuming errors on fitted diameter of circle (or length of major and minor axes of ellipse) to be equal to the resolution limit of AIA ($0.6\arcsec$/pixel) and IRIS ($0.166\arcsec$/pixel). Here, we clearly see the decrease in size of the oscillating region as we move down in the lower atmosphere as expected from the size of flux tube with height in the solar atmosphere. In the correlation images of AIA 304 \AA , IRIS 2796 \AA , AIA 1700 \AA , and HMI continuum, we also notice a shift of ($0\arcsec,0\arcsec$), ($1.5\arcsec,0.5\arcsec$), ($0\arcsec,0.6\arcsec$) and ($0\arcsec,1.2\arcsec$), respectively in the central position of black contour with respect to the coronal foot-point located in AIA 171 \AA\ passband. This shift can either be due to the inclination of the loop or due to the alignment errors between AIA, IRIS, and HMI. Moreover, we also notice that the identified loop foot-point at the photosphere is directed towards the umbral center rather than in any random direction with respect to the coronal foot-point of the loop. This shows that the loop is continuing towards the umbral center as visualized in Fig.~\ref{fig:map171}.

\begin{figure*}
    \centering
    \includegraphics[width=0.7\textwidth]{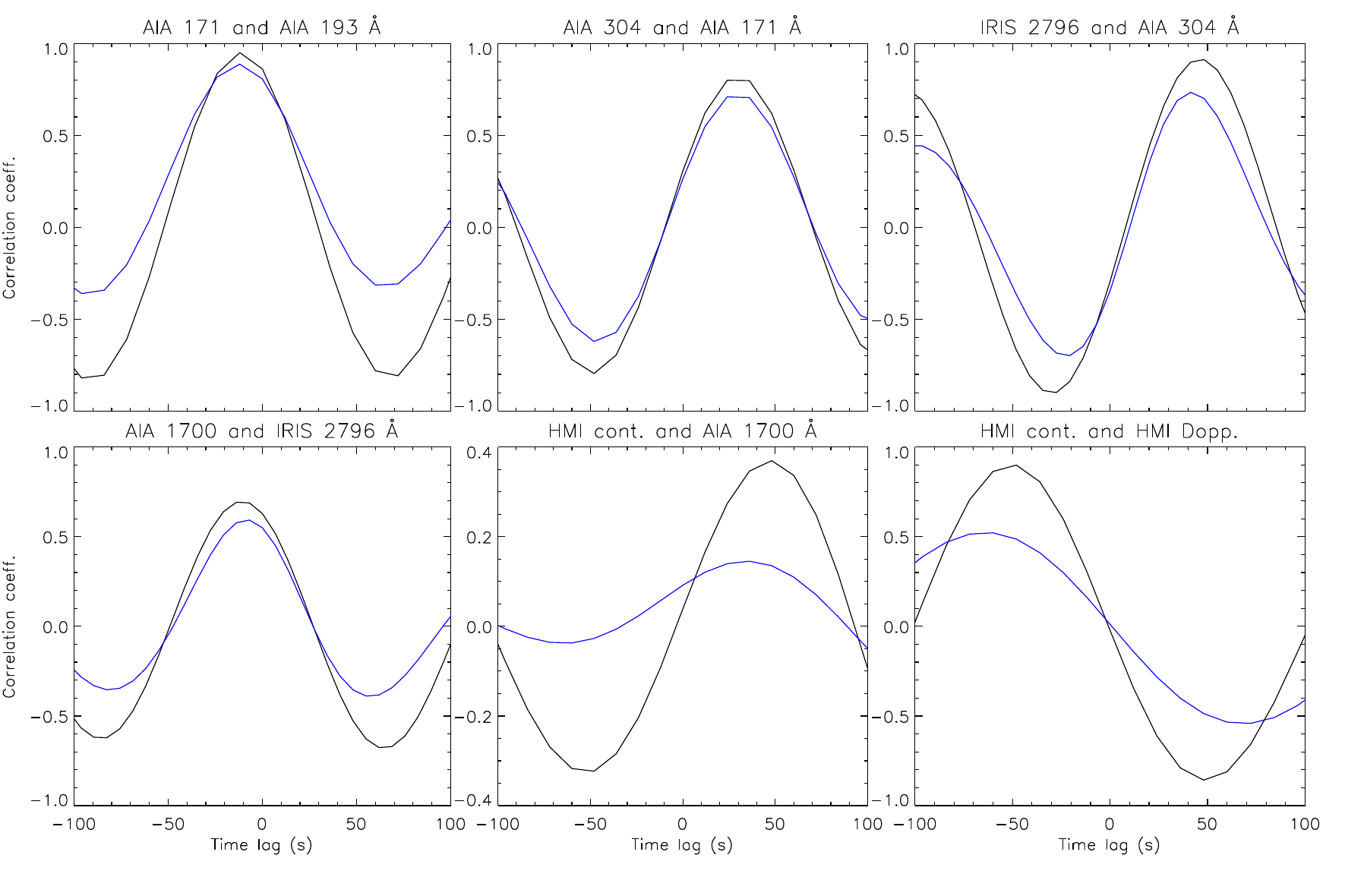}
    \caption{Variation of cross-correlation coefficients with respect to time lag obtained between light curves at loop 6 in different passbands as labelled. Black and blue lines show cross-correlation analysis performed on the Fourier-filtered and background subtracted light curves, respectively.}
    \label{fig:lag}
\end{figure*}

After determining the loop locations at each height, we obtained light curves at those locations from AIA, IRIS, and HMI passbands. We also performed, similar analysis using smaller time series, where we noticed an offset of $\pm$1 pixel for the maximum correlated pixel in Fig.~\ref{fig:alignment}. Therefore, we obtained light curves from $3\times 3$ pixel$^2$ binning for all the AIA and HMI passbands. This binning also incorporates alignment uncertainties of about 2 pixels in different AIA passbands \citep{2013ApJ...766..127Y}, and any lateral shift of loops from the photosphere to corona \citep{2015ApJ...812L..15K}. Similarly for IRIS, we are taking $5\times5$ pixel$^2$ binning. These pixel binnings also improve strength of the signals. In the left panels of Fig.~\ref{fig:lightcurve}, we plot original light curves obtained at loop locations at various atmospheric heights as labelled. Overplotted blue lines represent the background trends, which were obtained by taking the running average of 16-min. In the right panels, we plot 1-min smoothed background subtracted light curves. Detrended light curves provide clean intensity oscillations without any low-frequency background at all the atmospheric heights.

\subsection{Fourier analysis and filtration}

\begin{figure*}
    \centering
    \includegraphics[width=0.4\textwidth,angle=90]{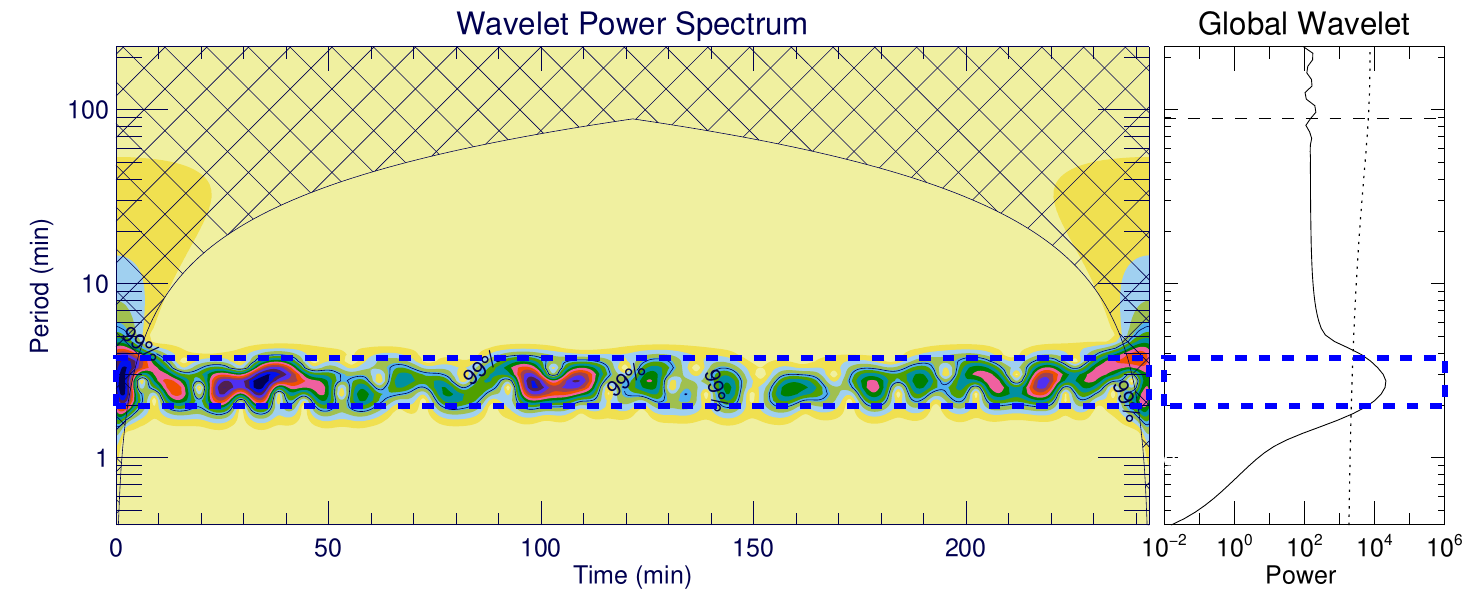}
    \caption{The left panel shows colour-inverted wavelet power spectrum of filtered light curve of loop 6 coronal foot-point obtained from AIA 171 \AA\ passband with 99\% confidence-level contours. Region marked with crossed lines denote cone-of-influence. The right panel shows the global wavelet power spectrum obtained by averaging wavelet power spectrum over time. The thin dashed lines specify the maximum period detectable from wavelet analysis because of cone-of-influence, and the dotted line specifies 99\%  confidence level curve. The overplotted blue colour rectangular boxes denote period window extracted for detail analysis. The time axis starts at 7:19 UT.}
    \label{fig:wavelet}
\end{figure*}

\begin{table*}
    \centering
        \caption{Correlation values obtained from both background subtracted and Fourier-filtered light curves for loop 6 between different pairs of atmospheric heights. Wave propagation speeds at different heights are also provided.}
  \label{tab:1} 
  \resizebox{\textwidth}{!}{
    \begin{tabular}{|c|c|c|c|c|c|c|c|} 
    \hline
\multicolumn{1}{|c|}{Passbands (\AA)}   &  \multicolumn{1}{|c|}{Distance (km)} &
\multicolumn{2}{|c|}{Correlation with background subtraction }  &
\multicolumn{2}{|c|}{Correlation with Fourier filteration} &
\multicolumn{1}{|c|}{Adiabatic acoustic} &  \multicolumn{1}{|c|}{Obs. speed (km s$^{-1}$)} \\ 
&  & Correl. coeff. & Time lag (s) & Correl. coeff. & Time lag (s) &  speed (km s$^{-1}$) \\
\hline
193-171    &   --      & 0.89 & $-12.0\pm12.0$  & 0.95 & $-12.0\pm12.0$ &  --  &  --              \\ \hline
171-304    & $102\pm78.5$  & 0.71 & $24.0\pm12.0$   & 0.80 & $24.0\pm12.0$  & 81.9$\pm13.5$    &  $4.2\pm3.9$  \\ \hline
304-2796   & $158\pm69$& 0.73 & $39.3\pm6.9$    & 0.91 & $46.2\pm6.9$   & 23.8 $\pm3.6$    &  $3.4\pm1.6$   \\ \hline
2796-1700  & $695\pm46$& 0.59 & $-8.9\pm6.9$    & 0.69 & $-15.8\pm6.9$  & 12.2 $\pm1.8$     &  $44\pm19.4$ \\ \hline
1700-cont. & $167\pm10$& 0.14 & $36.0\pm12.0$   & 0.37 & $48.0\pm12.0$  & 10.5 $\pm1.5$     &  $3.5\pm0.9$    \\ \hline
Dopp.-cont.& $62\pm5$  & 0.52 & $-60.0\pm12.0$  & 0.90 & $-48.0\pm12.0$ & 10.5 $\pm1.5$     &  $1.3\pm0.3$  \\ \hline
 \end{tabular}}
\end{table*}

In the left panels of Fig.~\ref{fig:filtered}, we show the Fast Fourier Transform (FFT) power spectrum obtained from original light curves of loop locations at various atmospheric heights as labelled. The FFT power spectrum is obtained using standard IDL routine $fft\_powerspectrum$. Plots show a wide distribution of power peaks in the period range of approximately 2-3.74 min at all the atmospheric heights, and also in the range of 4.2-6 min only at the photosphere. The two dominant period ranges are called 3-min and 5-min period bands, and are shaded with light and dark blue colours, respectively, in the left panels of Fig.~\ref{fig:filtered}. There are several nearby power peaks present in both the period bands. Vertical dashed lines represent the dominant peak in the 3-min and 5-min period bands which are present at all the atmospheric heights and at the photosphere respectively. We have ignored 3.74-4.2 min period window, which may have arised because of an artefact due to pixel crossing time due to differential rotation of sunspot \citep{2021RSPTA.37900175N}. Details of this artefact is described in Appendix~\ref{append:artefact}.

3-min period waves are generally observed in all the layers of the upper sunspot atmosphere \citep{2015LRSP...12....6K}. Sometimes they are also detected at the photospheric heights though weaker than the 5-min period \citep{2000ApJ...534..989B}. Since we have found significant power in the 3-min period band at every atmospheric height, we will utilise it for further detailed analysis. We apply a bandpass filter over the period range of 2-3.74 min (light blue colour shaded region in Fig.~\ref{fig:filtered}) on the original signals. The filtered signals are shown in the right panel of Fig.~\ref{fig:filtered}. Filtration provides a clean intensity oscillation in the 2-3.74 min period range. Due to several nearby power peaks in this period band, we see oscillations in the form of unclean wave packets. Such wave packets are also reported at several locations along the coronal fan loops  by \citet{2020A&A...638A...6S}.

In Fig.~\ref{fig:lag}, we plot cross-correlation coefficients as a function of time lag obtained between pairs of detrended (right panel of Fig.~\ref{fig:lightcurve}) and 3-min Fourier-filtered light curves (right panel of Fig.~\ref{fig:filtered}) as labelled with blue and black lines, respectively. Time lags obtained from both the light curves are consistent with each other except between AIA 1700 \AA\ and HMI continuum pair, where the correlation is very poor due to the presence of 5-min oscillations in the background subtracted light curve of HMI continuum. Using time lags obtained from filtered light curves and formation heights of various passbands, we determined the propagation speed of these 3-min waves at various heights. These estimated speeds provide lower limits on propagation speeds as line-of-sight projection effects are also involved. We have assumed formation heights for these passbands from the umbral atmospheric model of \citet{2022FrASS...9.1118D} which they claimed to be in agreement with  \citet{1999ApJ...518..480F,2009ApJ...707..482F} and others. Formation heights for photosphere HMI continuum is at 38 $\pm 1.9$ km, HMI Dopplergram 6173.34 \AA\ at $100\pm5$ km, temperature minimum AIA 1700 \AA\ at $205.0\pm10.25$ km, chromosphere IRIS 2796 \AA\ at $900.0\pm45.0$ km, transition region AIA 304 \AA\ at $1058.0\pm52.9$ km, corona AIA 171 \AA\ $1160.0\pm58.0$ km, and AIA 193 \AA\ at $1190.0\pm59.5$ km, see details in Fig.~\ref{fig:height} in Appendix~\ref{append:temp}. Estimated wave speeds at various atmospheric heights are provided in Table~\ref{tab:1}. Moreover, within the error bars, propagation speeds obtained in the lower atmosphere from different atmospheric models such as \citet{1986ApJ...306..284M} and \citet{2015ApJ...811...87A} are almost similar. Estimated speeds are less than the adiabatic acoustic speed  $c_s= \sqrt{\gamma 2 k_b T/\mu m_p}$, where $\gamma=5/3$ is adiabatic index, $k_b$ is Boltzmann constant, T is temperature, $\mu=1.27 $ is mean molecular weight, $m_p$ mass of proton \citep[e.g.,][]{2004psci.book.....A} at that particular height. This confirms that these are propagating slow magnetoacoustic waves except at chromospheric height where speed becomes supersonic. Moreover, it should also be noted that effects of partial ionization in the lower atmosphere can make $\gamma$ as low as $\gtrsim 1$ \citep[section 4.1, ][]{2004psci.book.....A}. However, in this case also observed propagation speeds will remain mainly subsonic. For error estimate, we have assumed error in time as the cadence of each passband which is 12 s and 6.88 s for SDO and IRIS respectively. Since \citet{2022FrASS...9.1118D}  did not provide any method to calculate errors in the formation heights, we have assumed an overall made up error of 5\% in all the formation heights as we noted $\approx 5\%$ to be the maximum error in height quoted by them.

 \begin{figure*}
    \centering
    \includegraphics[width=0.9\textwidth]{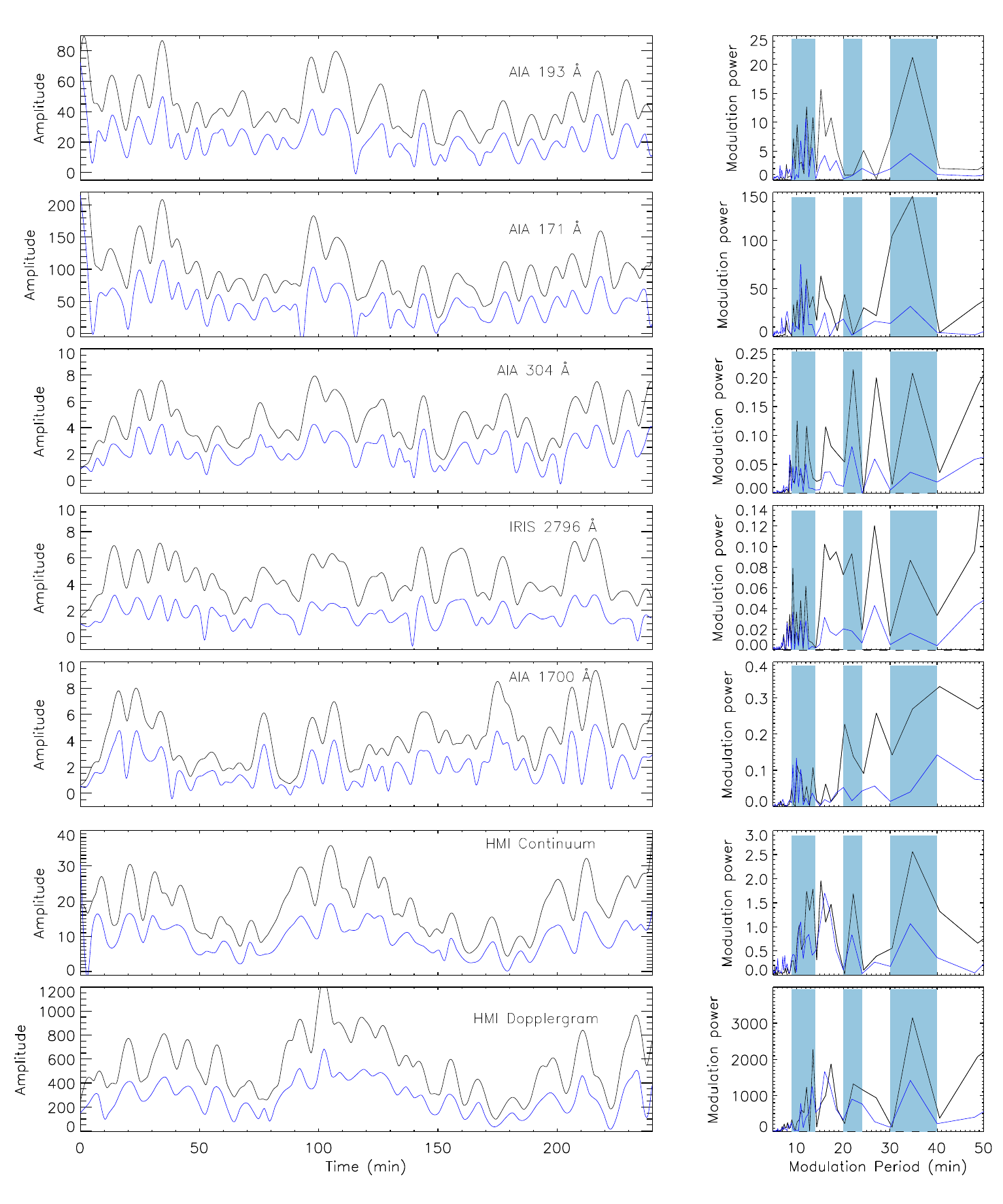}
    \caption{Left panels show the amplitude modulations of 3-min oscillations. Black and blue colour lines show amplitude modulations extracted from the wavelet spectrum and extrema (maxima) of all 3-min filtered light curves shown in Fig.~\ref{fig:filtered}, respectively. The right panels show the FFT power spectrum of corresponding amplitude modulations. Shaded regions in sky blue colour highlight dominant modulation periods observed at different atmospheric heights.}
    \label{fig:am}
\end{figure*}

\begin{figure*}
    \centering
    \includegraphics[width=0.7\textwidth]{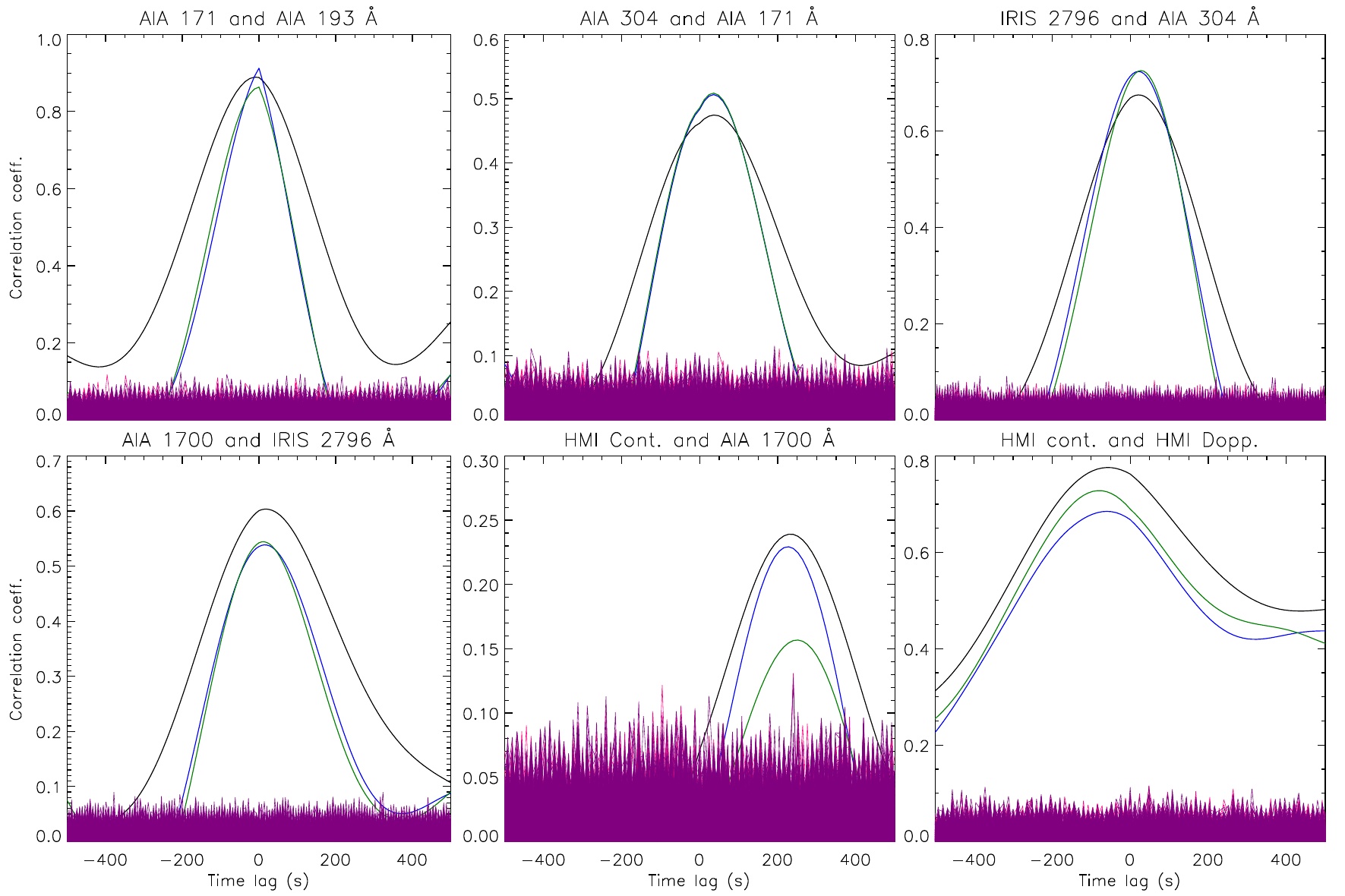}
    \caption{The black, green and blue coloured lines show correlation coefficients with respect to time lag between amplitude modulation curves obtained from wavelet, maxima and minima respectively at different atmospheric heights as labelled and shown in the left panel of Fig.~\ref{fig:am}. Pink and purple colour lines are obtained by carrying out correlation analysis on randomized amplitude modulation curves from wavelet and maxima curves respectively (see details in Appendix~\ref{append:random}).}
    \label{fig:co_am}
\end{figure*}

\subsection{Wavelet analysis}

The above Fourier analysis provided information in the period domain only. Therefore, wavelet analysis is incorporated to obtain the variation of power of these 3-min oscillations with time. Here, we are using the Morlet wavelet for our analysis and utilising the wavelet tool developed in IDL by \citet{1998BAMS...79...61T}. 

In Fig.~\ref{fig:wavelet}, we show the wavelet power spectrum of filtered light curve of coronal foot-point of loop 6 as obtained from AIA 171 \AA\ passband (see right panel of Fig.~\ref{fig:filtered}). The left panel shows the wavelet power spectrum with time on the x-axis and period on the y-axis, showing variation of oscillatory power. Different colour contours show varying power densities, with blue being the highest. Here, the region marked with crossed lines is called the Cone-of-Influence (COI) which refers to the region where the transform suffers from the edge effect. Oscillation periods in this region are unreliable. In the right panel of Fig.~\ref{fig:wavelet}, we show the global wavelet power spectrum, which is obtained by taking the average over the time domain of the wavelet transform. The thin dashed line shows the maximum period detectable from the wavelet analysis due to the COI. 

The overplotted blue rectangular box shows 2-3.74 min period window in the wavelet and global wavelet power spectrum. This period window will be utilised to extract more detailed properties of 3-min waves. We are extracting amplitude modulations of 3-min waves from this rectangular box, and describe them in the following subsections.

\subsection{Modulations}
\label{section:all}

In the left panel of Fig.~\ref{fig:filtered}, we can see many closely spaced frequencies in the 3-min period band, which results in amplitude modulation of the signal as shown in the right panel of Fig.~\ref{fig:filtered}. In Fig.~\ref{fig:wavelet}, we can see that the power is not constant but changes with time. Using this property of 3-min waves, we explore the source region of these waves in the lower atmosphere, and thus the magnetic connectivity of the solar atmosphere.

\subsubsection{Amplitude modulation}

 To determine the modulation of the amplitude of these 3-min waves, we extracted the wavelet power in 2-3.74 min window as shown by the blue box in the left panel of Fig.~\ref{fig:wavelet}. We then averaged the extracted window over period, and thus obtained amplitude variations (square root of wavelet power) with time for all the light curves at different atmospheric heights. We also devised another method in which we have identified the local maxima and minima of the filtered light curves using IDL routine $EXTREMA$ to trace the amplitude modulation envelopes. Obtained amplitude modulations with time from wavelet and extrema (-maxima) methods are plotted in the left panels of Fig.~\ref{fig:am} with black and blue colour lines, respectively,  for all the atmospheric heights as labelled. We cross-correlated all the amplitude modulations obtained at different atmospheric heights with the nearest atmospheric layers. Amplitude modulation correlations obtained from wavelet, and extrema -maxima and -minima are plotted in black, green and blue colour lines, respectively, in Fig.~\ref{fig:co_am}.  Time lags  obtained from the wavelet and extrema (-maxima and -minima)  methods  match well with the time lags obtained in Fig.~\ref{fig:lag} within the error bar. However, with exception for HMI-continuum and AIA 1700 \AA\ pair where the time lag of amplitude modulation matches well with the time lag of second peak of the 3-min correlation curve at 228 s, which has a higher correlation value compared to the first peak at 48 s. Moreover, correlation value decreases if we correlate the amplitude modulation curves from neighboring pixels, and thus supports the robustness of our loop tracing. Correlated 3-min amplitude modulations at different atmospheric heights clearly indicate that these waves originate in the lower atmosphere, and different atmospheric heights are coupled together. Thus, it provides clear evidence of magnetic connectivity of the whole solar atmosphere.

\begin{figure*}
    \centering
    \includegraphics[width=0.75\textwidth]{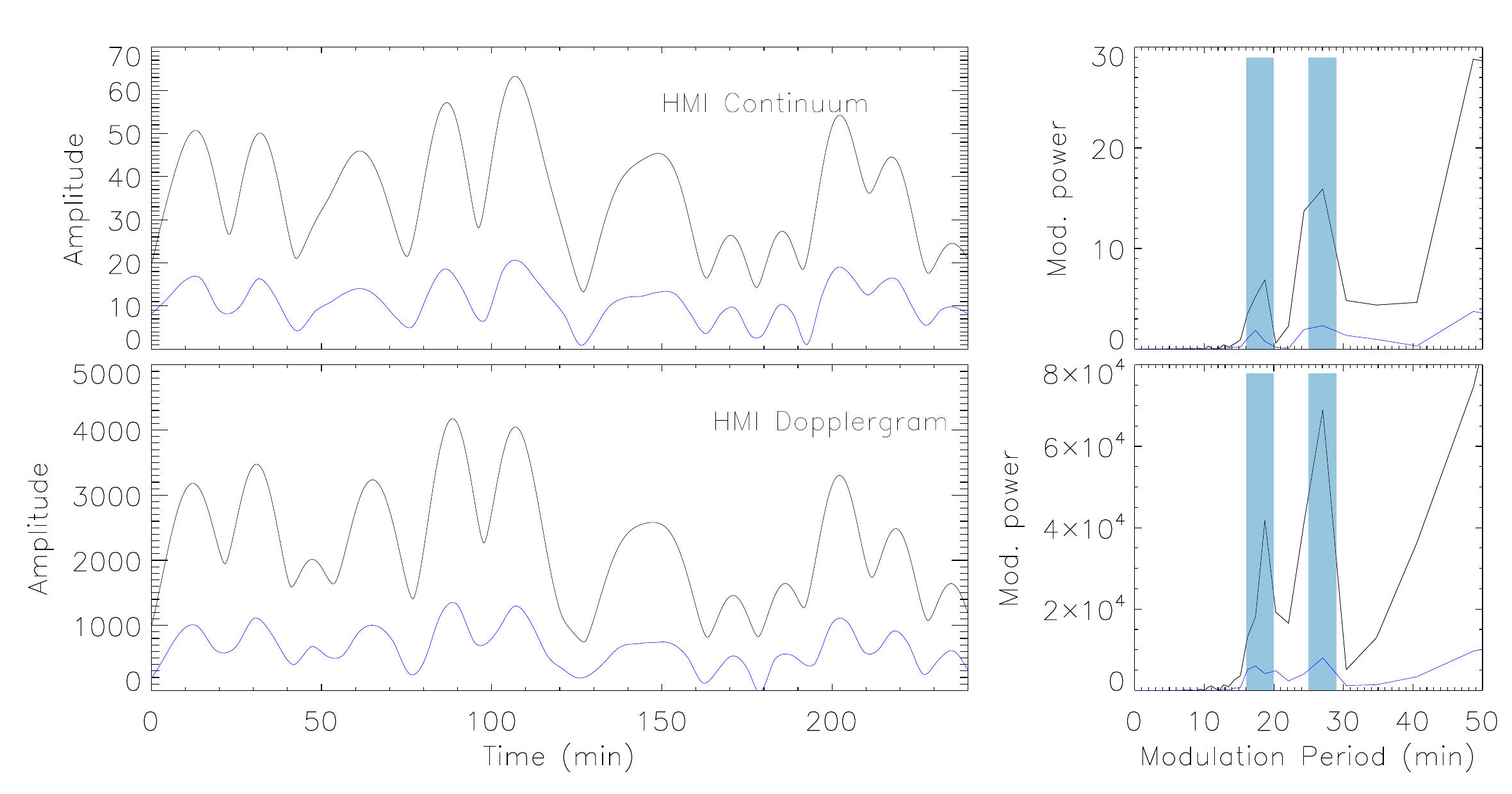}
    \caption{The left panels show the amplitude modulation of 5-min oscillations on the photosphere (HMI continuum and Dopplergram) as labelled. Black and blue colour lines show modulations extracted from the wavelet spectrum and extrema of 5-min filtered light curves, respectively. The right panels show the FFT power spectrum of corresponding modulations. Shaded regions in sky blue colour highlight dominant modulation periods observed in different HMI passbands.}
    \label{fig:photosphere_period}
\end{figure*}

 Further, to check the periods of amplitude modulation, we obtained the FFT power spectrum of amplitude modulations, and plotted them in the right panels of Fig.~\ref{fig:am}. Here we can clearly see that dominant amplitude modulation periods are approximately in the range of  9--14 min, 20--24 min, and 30--40 min and are present at all the layers of the solar atmosphere as shown by shaded regions in the right panels of Fig.~\ref{fig:am}.  These modulations are formed due to the various combinations of closely spaced power peaks present within the 3-min period band. Some of the dominant periods present within this band are 2.51 min (6.64 mHz), 2.7 min (6.17 mHz), 2.8 min (5.95 mHz), 3.08 min (5.41 mHz), etc. Combinations of these periods like  2.51 and 2.7 min,  2.7 and 3.08 min, 2.8 and 3.08 min, 2.7 and 2.51 min etc. can result in beat periods of 35.67 min, 21.88 min, 30.80 min, 10.59 min, etc. respectively. These derived beat periods are within the range of observed modulation periods as noted above at all the layers of solar atmosphere. Presence of similar modulation periods at all the heights signify that modulations of 3-min waves are essentially coupled to each other at different atmospheric layers, as also found from the correlation analysis. Thus, the result provides clear evidence of upward propagating waves from the photosphere to the corona along the observed fan loop at different atmospheric heights.   
 
\subsubsection{Relation between modulations of 3-min and 5-min oscillations at the photosphere}

\begin{figure}
    \centering
    \includegraphics[width=0.49\textwidth]{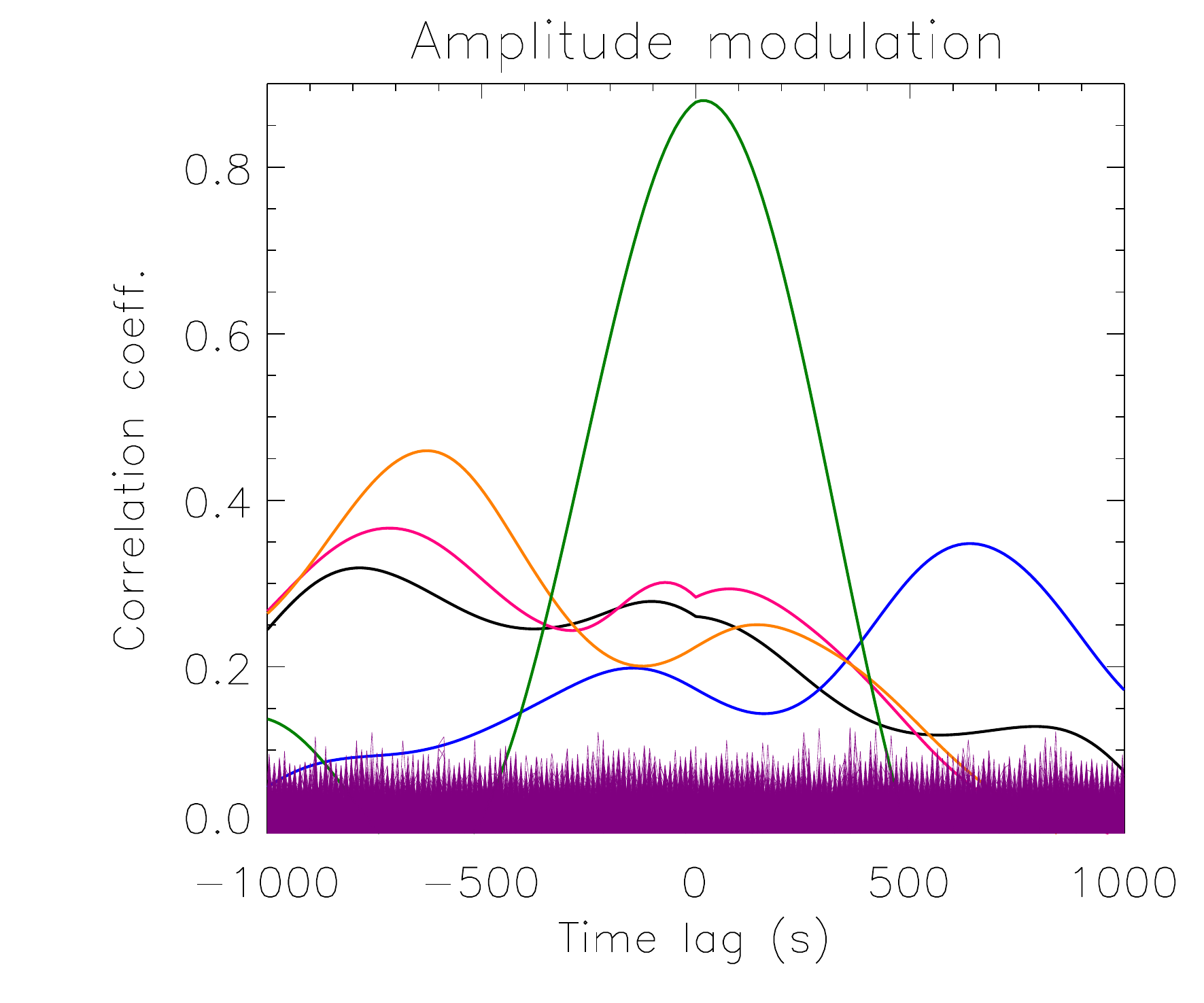}
    \caption{Correlations between amplitude modulation curves of 3-min and 5-min oscillations obtained from maxima with respect to time lag. Orange, black, pink, blue, and green colour lines show correlations obtained between Fourier filtered curves of 3-min and 5-min period windows obtained from HMI Dopplergram, 3-min and 5-min period windows obtained from HMI continuum, 3-min window of continuum and 5-min window of Dopplergram, 5-min window of continuum and 3-min window of Dopplergram, and 5-min window of continuum and 5-min window of Dopplergram, respectively. Purple colour lines are obtained by carrying out correlation analysis on respective randomized modulation curves (see details in Appendix~\ref{append:random}).}
    \label{fig:photosphere}
\end{figure}

\begin{table*}
    \caption{Amplitude modulation periods (min) of 3-min oscillations observed in various fan loop locations at different atmospheric heights as marked in Figs.~\ref{fig:maps}, \ref{fig:map171}.}
     \label{tab:ama}
   \centering
    \begin{tabular}{|c|r|r|r|r|r|r|r|r|} 
    \hline
Passbands (\AA)  & loop 1 & loop 2 & loop 3 & loop 4 & loop 5 & loop 6 & loop 7 & loop 8 \\
\hline
AIA 193 & 23,30 & 12,15,27,35 & 15,30 & 13,16,24 & 14,19,27,40 &12,15,35 & 13,18,40 & 14,20,27,35 \\ \hline
AIA 171 & 12,22,40 & 12,15,27,35 & 16,30 & 13,16,24 & 14,19,27 & 12,21,35 & 13,18,40 & 15,20,27,35\\ \hline
AIA 304 &  14,20,40 & 16,21,31 & 16,30 & 13,16,24,34 & 14,18,30,40 & 12,22,27,34 & 14,27,40 & 16,20,27,35 \\ \hline
IRIS 2796 & 12,19,30,40 & 16,20,34 & 12,16,31 & 13,20,34& 10,22,27 & 12,22,27,34 & 14,20,34 & 16,20,35\\ \hline
AIA 1700 & 13,16,20,24 & 16,22,35 & 16,22,35& 14,17,35 &10,22,40 & 13,20,27,40 & 14,20,37 & 16,20,35\\ \hline
HMI continuum & 12,21,25,35 &15,27,35 & 15,27,35 & 17,22,30 & 22,40 & 13,22,35 & 13,27 & 12,20,35\\ \hline
HMI Dopplergram & 14,21,40 & 15,22,27,40 & 16,27,40 & 17,22,30 & 20,40 & 13,22,35 & 13,30 & 11,20,35\\ \hline
    \end{tabular}
\end{table*}

\begin{table*}
    \caption{Amplitude modulation periods (min) of 5-min oscillations observed in various fan loop locations at the photosphere as marked in Figs.~\ref{fig:maps}, \ref{fig:map171}.}
     \label{tab:ama5}
   \centering
    \begin{tabular}{|c|r|r|r|r|r|r|r|r|} 
    \hline
Passbands (\AA)  & loop 1 & loop 2 & loop 3 & loop 4 & loop 5 & loop 6 & loop 7 & loop 8 \\
\hline
HMI continuum & 23,41 &22,35 & 17,35 & 20,30 & 23,40 & 18,27 & 27,50 & 27,48\\ \hline
HMI Dopplergram & 17,23,41 & 22,41 & 17,22,40 & 20,35 & 23,48 & 17,27 & 27,50 & 27,48\\ \hline
    \end{tabular}
\end{table*}

As evident from the FFT power spectrum obtained from HMI continuum and Dopplergram curves in Fig.~\ref{fig:filtered}, both 3-min and 5-min oscillations are present at the photosphere. Therefore, we also obtain amplitude modulation of 5-min oscillation at the photosphere to explore any relation between both the oscillations. In Fig.~\ref{fig:photosphere_period}, we plot amplitude modulation of 5-min oscillation as labelled, and their respective FFT power spectrum. From the plots, we can clearly see that dominant amplitude modulation periods are in the range of 16--20, and 25--29 min as shown by shaded regions in the right panels. On comparing it with modulations of 3-min oscillations in Fig.~\ref{fig:am}, we notice that one of the dominant short period ($\approx 12$ min) modulations for 3-min oscillations present at all heights is not present in the modulations of 5-min oscillation. However, other longer periods, such as 20 min band and above, are present in both modulations. This indicates that 5-min oscillations are unable to capture small time-scale modulations as found in the modulations of 3-min oscillations. We also noticed that 27 min period detected in 5-min amplitude modulation at the photosphere is not present in the 3-min amplitude modulation at the photosphere. However, it can be observed at the chromosphere (AIA 1700 \AA\ and IRIS 2796 \AA) and transition region (AIA 304 \AA). This may indicate that amplitude modulations observed in 5-min oscillations are also propagating upward through 3-min oscillations. Therefore, we further carry out correlation analysis to verify any coupling between these modulations.

In Fig.~\ref{fig:photosphere}, we plot the correlation coefficient with respect to time lag obtained between amplitude modulations of 3-min and 5-min oscillations observed from HMI continuum and Dopplergram as labelled. Maximum correlation coefficient between amplitude modulation of 5-min and 3-min oscillations from HMI Dopplergram is about 0.46, with a time lag of -624 s. Similarly, the corresponding amplitude modulation correlation obtained from the HMI continuum is about 0.35 with a time lag of -780 s. Although these correlation values are 
reasonable, large time lags obtained between 3-min and 5-min oscillations are unclear and demand a dedicated investigation. 

In Fig.~\ref{fig:co_am}, we can clearly see that correlation values are above the error range provided by randomization test (see details in Appendix~\ref{append:random}). This implies that amplitude modulation time series of upper height is clearly correlated to time series of lower height. Hence, we can say that 3-min amplitude modulation originated at the photosphere and is moving upward as waves move upward. Furthermore, we also carried out analysis as described in \citet{2015ApJ...812L..15K} by taking the same pixel-to-pixel correspondence of the loop locations at different heights. Although our findings remain similar, correlation values obtained between the light curves and amplitude modulation curves at different heights were less than the our maximum correlation values. This is expected as we are maximizing our correlation by shifting the loop location to the maximum correlated pixel location in the neighborhood at different heights.

We performed a similar analysis on the other seven fan loops identified in Fig.~\ref{fig:map171}. Here also loop foot-points at the photosphere are directed towards the umbral center, and their oscillating regions are smaller with respect to their coronal counterpart as found for loop 6. Estimated wave propagation speeds and correlation coefficients at various atmospheric heights for all the loops are provided in Tables~\ref{tab:append1} and \ref{tab:append2} in Appendix~\ref{append:wavespeed}. Estimated speeds are less than the acoustic speed at that particular height which confirms that these are propagating slow magnetoacoustic waves except at chromospheric height as also found for loop 6 (see Table~\ref{tab:1}). Within the error bar, propagation speed increases with height which essentially reflects the increase in acoustic speed, and thus increase in temperature with height. However, these propagation speeds should be taken very cautiously due to the strong inclination effects along the loops as visible from Figs.~\ref{fig:maps} and \ref{fig:map171}. Summary of amplitude modulation periods of 3-min and 5-min oscillations at various atmospheric heights for all the loops are provided in Tables~\ref{tab:ama} and \ref{tab:ama5}. Obtained modulation periods are in the range of 10-40 min, similar to those in loop 6. Some of the common modulation periods found in all the fan loops are in the range of 12-14, 20-22, 27-30, and 35-40 min. All the fan loops show at least one common modulation period which is observed at all the atmospheric heights. We also found similar pattern in correlation of modulations between various atmospheric heights as for loop 6. Correlations become poor at heights where atmospheric parameters change drastically, as also found for loop 6. Obtained results suggest more or less similar statistics near the foot-points of these loops at various atmospheric heights. Such similarity in behaviour of fan loops emanating from the same sunspot umbra is expected as also noted by \citet{2009SSRv..149...65D}. Obtained results confirm that 3-min waves observed in coronal fan loops originate at the photosphere and are propagating upward as they have at least one common modulation period at each atmospheric height \citep[e.g.,][]{2015ApJ...812L..15K}. These results again provide clear evidence of magnetic connectivity of the whole solar atmosphere.

\section{Discussion and Summary}
\label{sec:discussion}

In this work, we traced coronal fan loops from the corona to the photosphere in the sunspot umbra. \citet{2012ApJ...757..160J} found that  coronal fan loops are anchored at the photosphere where umbral dots are located. \citet{2019ApJ...873...75S} also noticed that bright loops rooted in the sunspot are result of light bridge and umbral dots activities. However in our observation, due to the limited spatial resolution of the photospheric dataset (HMI continuum), we could not verify whether identified photospheric foot-points of fan loops are umbral dots as diameter of these dots varies from $0.2\arcsec$ to $0.8\arcsec$ \citep{2012ApJ...757..160J,2018SoPh..293...54Y}. 

We also obtained the area of the oscillating region at various atmospheric heights using the contour levels. We surmise that these areas may translate to cross-sectional areas of fan loops at different heights (see Fig.~\ref{fig:alignment}). However, caution must be taken while selecting contour levels which varies among different loops. This demands a detailed investigation to translate the area of oscillating region to cross-sectional area of the analysed loop. Here, area of  oscillating region decreased from $\approx 2.43\pm1.17$ arcsec$^2$ ($8.24\pm2.14$ arcsec$^2$) in the corona to $\approx 0.18\pm1.77$ arcsec$^2$ ($2.06\pm1.07$ arcsec$^2$) at the photosphere for 95\% (80\%) contour level. Expansion of the oscillating region which may be associated with expansion of flux tube with height is expected in the solar atmosphere because pressure inside and outside of flux tube changes differently with height \citep[see details in][]{2004psci.book.....A}. Such expansion of flux tube area from the photosphere to the chromosphere is well noted in the magnetic bright points by \citet{2009Sci...323.1582J} and in umbral dots from the photosphere to chromosphere by \citet{2012ApJ...757..160J} using the power maps of 3-min oscillations. However, here we find clear evidence of such expansion of coherent oscillating regions from the photosphere to corona as described in Sec.~\ref{sec:correlation} which were not reported before.

 Although wave speeds are estimated in umbral atmosphere at few heights \citep[e.g.,][]{2002A&A...387..642O,2012ApJ...746..119R,2013A&A...554A.146K}, here we estimated the propagation speed of waves from the photosphere to the corona along the identified fan loop locations. Measured speeds are less than adiabatic acoustic speed at those heights. This confirmed that these are propagating slow magnetoacoustic waves. Moreover, we also noted that the time lags obtained in Fig.~\ref{fig:lag} are averaged over multiple frequencies present in the light curves over the time duration of 4-hours. Further analysis reveals that some of these delays are frequency as well as time-interval dependent. Similar features were also noted by \citet{2012ApJ...746..119R}. However, here we are utilising only the averaged time lag to calculate the wave propagation speed. Detailed analysis of time interval and frequency-dependent time lags (phase delays) between the oscillations at different heights will provide information on the dispersive nature of the medium (flux tube) and will be of future interest. From Table.~\ref{tab:1}, we notice that time lag between AIA 171 and 193 \AA\ passbands is negative (with equal error) even though AIA 193 \AA\ passband is sensitive to higher temperature and height \citep[e.g.,][]{2012ApJ...746..119R}. Since AIA passbands are multithermal in nature, AIA 193 \AA\ passband has some contributions from cooler O V and Fe VII lines which detect plasma from transition region temperatures \citep{2012SoPh..279..427K}. As temperature of fan loops are found to be around 0.9 MK \citep[e.g.,][]{2017ApJ...835..244G}, therefore oscillations recorded in AIA 193 \AA\ passband could be due to the cooler plasma formed at lower height and temperature than the AIA 171 \AA\ passband. This can explain the negative time lag obtained between the two passbands. Average time lag between chromospheric and photospheric oscillations is also negative, which may indicate some downward propagation of waves. This may also suggest that at this height, some of the magnetoacoustic waves are reflected back \citep{2006ApJ...653..739K}. Simulation result of \citet{2013A&A...551A.137M} for flux tube suggests the presence of both fast and slow wave modes superimposed on the same structure.  Recent result of \citet{2023ApJ...944L..52C} also shows downward propagation of chromospheric oscillations in the sunspot umbra, and thus in agreement with our findings.
 
To probe the source region of 3-min waves along the fan loops, and thus the connectivity of the solar atmosphere, we utilised amplitude modulations of these 3-min waves. Amplitude modulations of 3-min waves are already noted in the chromosphere \citep[e.g.,][]{2006ApJ...640.1153C}, transition region \citep[e.g.,][]{2006ApJ...643..540M}, and corona \citep[e.g.,][]{2020A&A...638A...6S}. \citet{2015ApJ...812L..15K} found almost similar periods of amplitude modulations from the photosphere to the corona, from which they concluded that photospheric p-modes externally drive 3-min slow magnetoacoustic waves observed in the coronal fan loops. However, their conclusion was only based on the similar periodicity of amplitude modulations of the order of 20-27 minutes across all the observed passbands covering the photosphere to coronal heights. Moreover, they compared the modulation period of 5-min oscillation at the photosphere with the modulation period of 3-min oscillations in the upper atmosphere to establish the source of 3-min waves. However contrary to this, \citet{2006ApJ...640.1153C} inferred that power in the 3-min oscillations observed at the chromospheric heights is directly related to photospheric 3-min oscillations by means of linear wave propagation, rather than any nonlinear interaction of 5-min waves. In this work, we found presence of 3-min oscillations at the photospheric foot-point of the fan loops. Such 3-min oscillations at the base of fan loops were also observed by \citet{2012ApJ...757..160J} at the photosphere. \citet{2012A&A...539L...4S} also found a clear presence of 3-min waves at the photosphere, which were strictly confined to the umbral region.  Moreover, we found several correlated amplitude modulation period ranges for 3-min waves such as 9-14 min, 20-24 min, and 30-40 min (see Fig.~\ref{fig:am}). However, a strong 27 min modulation period was also noted at some atmospheric heights. We have selected only those dominant modulation period ranges which are present at almost all the atmospheric heights. Therefore, based on our findings of similar modulation periods of 3-min waves at all the atmospheric heights, we conclude that 3-min waves in the upper atmosphere are direct result of 3-min oscillations observed at the photospheric umbral region.

Moreover, we also noticed a well known change in the dominant  period of oscillations with increasing height above the sunspot umbra, i.e. from  5-min in the photosphere to 3-min in the chromosphere \citep[e.g.,][]{2016PhDT........15L}. There are two popular mechanisms proposed to explain this period shift, chromospheric acoustic resonator \citep{1983SoPh...82..369Z}, and linear propagation of waves with frequencies above the acoustic cutoff from the base of photosphere \citep{1991A&A...250..235F,2006ApJ...640.1153C}. Recently, \citet{2020NatAs...4..220J} demonstrated the presence of chromospheric resonance cavity above the sunspot, however, see also \citet{2021NatAs...5....2F}. \citet{1991A&A...250..235F} in their model surmised that 3-min oscillations are excited already at the base of photosphere by 5-min oscillations. Recent numerical simulation of \citet{2019A&A...627A.169F} concluded that both mechanisms require excitation of 3-min waves at the lower heights.  Recent models of \citet{2018MNRAS.481..262W} and \citet{2019A&A...623A..62K} reported that 5-min driving acoustic waves can excite oscillations of shorter periods due to wave reflection. In this work, we have found presence of 3-min oscillations along with dominant 5-min oscillations at the photosphere. Based on these findings and above reported results, we can expect that some of the observed properties of 3-min oscillations at the photosphere, chromosphere, and above will be similar to that of 5-min oscillations at the photosphere due to their common origin. This explains some similarities found between modulation periods of 3-min oscillations and 5-min oscillations in our analysis. Henceforth, these findings in the lower atmosphere are in accordance with the results of \citet{1991A&A...250..235F} and \citet{2006ApJ...640.1153C}.

 On the basis of a similar modulation period of 5-min and 3-min oscillations, \citet{2015ApJ...812L..15K} concluded photospheric p-modes as the driver of 3-min waves in coronal fan loops. However, \citet{2017ApJ...836...18C} concluded that 3-min and 5-min oscillations are not connected and occur independently. Furthermore, simulations of \citet{2020ApJ...892L..31C} suggest that 3-min and 5-min oscillations are coupled and originate from the same source, which lies 1000-2000 km below the photosphere. Whereas \citet{2008SoPh..248..395S} and \citet{2023MNRAS.519.4397S} found that 3-min oscillations are localised in the umbra while 5-min is present at the umbra-penumbra boundary. In this work, we also explored coupling between 3-min and 5-min oscillations at the photosphere using HMI continuum and Dopplergram data. We found a reasonable correlation between modulations of 3-min and 5-min oscillations with huge time lags, as shown in Fig.~\ref{fig:photosphere}. We also examined the modulation periods of both the oscillations and found the $\approx20$ min period to be common in amplitude modulation of 3-min and 5-min oscillations. However, 12 min modulation period observed in 3-min oscillations is not present in the modulation of 5-min oscillations. Furthermore, simulation result of \citet{2020ApJ...892L..31C} assumes a non-dispersive medium, thus can not explain the huge time lag obtained between 3-min and 5-min oscillations in this observation. Therefore, any connection between 3-min and 5-min oscillations at the photosphere is still unclear, and demands a dedicated study to explore their connection in detail.

Source region of coronal 3-min waves at the photosphere showed enhancement of power of 3-min oscillations in the umbral dots where foot-points of coronal fan loops were anchored \citep{2012ApJ...757..160J}. Similarly, \citet{2013A&A...554A.146K}  also found concentration of high frequency ($\approx 2-min)$ oscillations at the temperature minimum and chromospheric levels, and in small regions inside the umbra at the photospheric level, which they speculated to be the location of umbral dots. Furthermore, results of \citet{2017ApJ...836...18C} also suggests that the 3-min chromospheric oscillations in the sunspots are generated by 3-min oscillation occurring in the light bridges and umbral dots due to magneto-convection. However, as noted before, due to the limited spatial resolution of the current dataset (HMI continuum), we could not verify whether identified fan loop foot-points are anchored in the umbral dots.
 
We also examined the correlations in modulations at different heights (see Figs.~\ref{fig:co_am}), and found that the solar atmosphere is coupled. However, we also noticed that correlation becomes poor between some height pairs (HMI continuum \& AIA 1700 \AA , and AIA 304 \& AIA 171 \AA), i.e. at heights where a sharp transition in density or temperature occurs. One possible reason for this decrease could be the amplitude steepening of the waves due to drastic change in density which leads to shock formation, especially in the chromosphere and transition region \citep{2000SoPh..192..373B}. Period of these shocks is also found to be around 3-min. 

Furthermore, it should also be noted that finding the connectivity in the solar atmosphere through wave propagation is a very complex task due to drastic change in the dynamics of photosphere, chromosphere and corona. Umbral photosphere is a complex region due to the presence of plasma-$\beta= 1$ layer close to the formation height of the radiation which affects the wave modes \citep[e.g,][]{2015ApJ...807...20P,2020ApJ...900L..29F}. Propagation and visibility of waves in the photosphere and chromosphere are also affected by opacity effects \citep[e.g.,][]{2000ApJ...534..989B,2003ApJ...588..606K,2014ApJ...795....9F}. Therefore, for a better understanding of propagation of waves in the solar atmosphere, detailed MHD simulations with non-local thermodynamic equilibrium (NLTE) forward modeling will be required in order to interpret the observations of different wave modes and other dynamical processes at various atmospheric heights \citep[e.g.,][]{1992ApJ...397L..59C,2023A&A...670A.133F}.

 In summary, we probed the magnetic coupling of the solar atmosphere using amplitude modulation of 3-min waves observed from the photosphere to the corona. The source region of propagating 3-min slow magnetoacoustic waves observed in the coronal fan loops was traced down into the umbral region of the photosphere. These 3-min waves showed periodic modulations in amplitude with periods in the range of about 9-14 min, 20-24 min and 30-40 min, and are correlated at different atmospheric heights. Results reveal that 3-min waves observed in the coronal fan loops are driven by 3-min oscillations observed at the photospheric foot-points of these fan loops  within the umbra which may be the sites of umbral dots. Result provides clear evidence of wave propagation from the photosphere to the corona through chromosphere, and transition region. These results can provide helpful insights in the modeling of wave propagation in the solar atmosphere. Furthermore, we found reasonable correlation between the 3-min and 5-min oscillations at the photospheric foot-points in the umbra. However, to conclude anything convincingly between both oscillations, a dedicated long duration high-resolution observations are needed so as to resolve the umbral dots to perform detailed analysis. This may shed some light on whether the 3-min oscillations are driven independently by magneto-convection or are driven by the 5-min p-modes.
   
\section*{Acknowledgments}
We thank the referee for helpful comments and suggestions that improved the quality of presentation. AIA and HMI data are courtesy of SDO (NASA). IRIS is a NASA small explorer mission developed and operated by LMSAL with mission operations executed at NASA Ames Research center and major contributions to downlink communications funded by the Norwegian Space Center (NSC, Norway) through an ESA PRODEX contract. Facilities: SDO (AIA, HMI). 

\section*{Data Availability}

The observational data utilized in this study from AIA and HMI on-board SDO are available at   \url{http://jsoc.stanford.edu/ajax/lookdata.html} and data from IRIS mission are available at \url{https://www.lmsal.com/hek/hcr?cmd=view-recent-events\&instrument=iris}.






\appendix

\section{Cross-section of coronal fan loop in AIA 171 \AA\ image}
\label{append:area}

\begin{figure}
    \centering
    \includegraphics[width=0.45\textwidth]{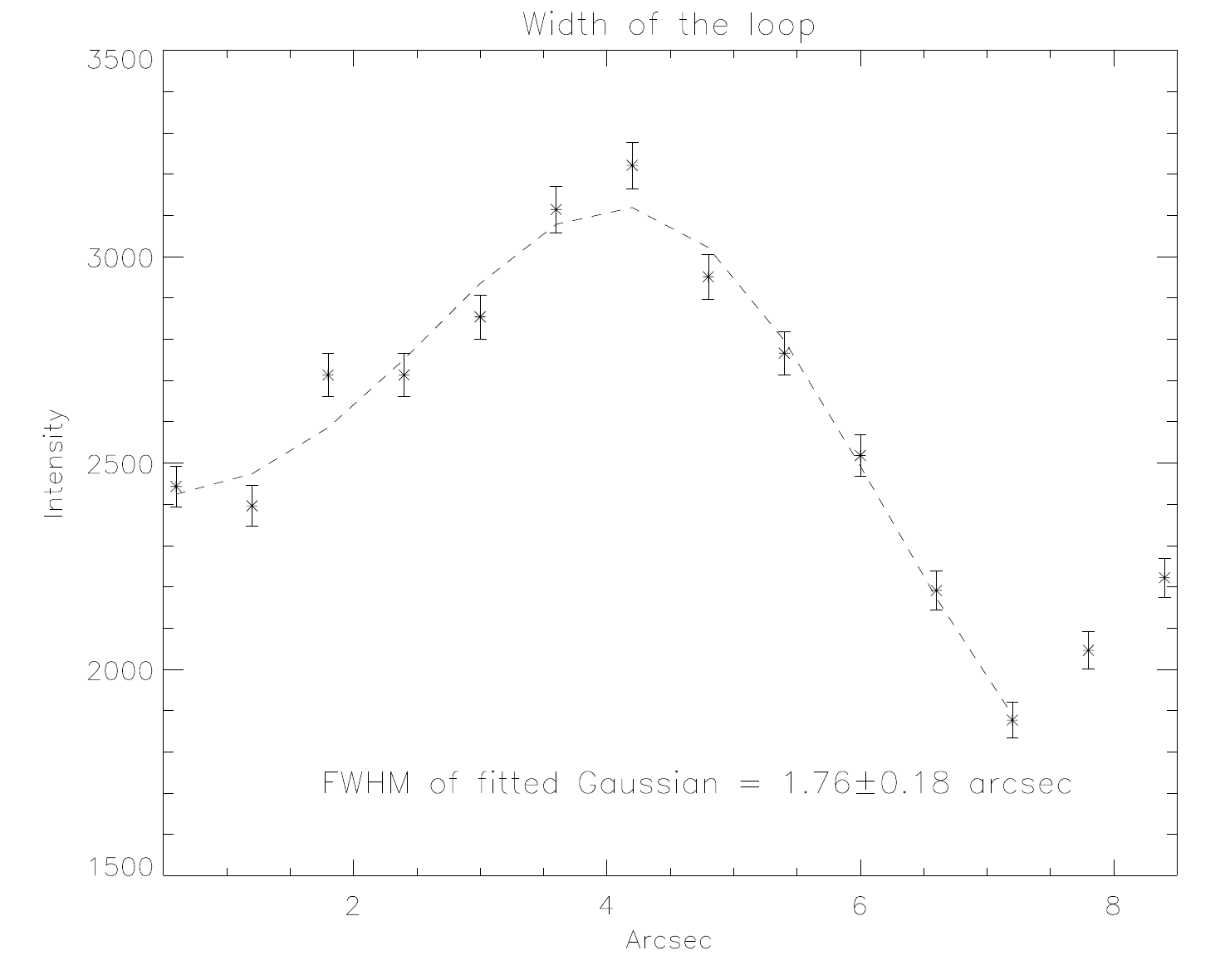}
    \caption{Gaussian with linear function fitted over the intensity distribution across the coronal foot-points of loop 6 as observed in AIA 171 \AA\ passband.}
    \label{fig:coronal_diameter}
\end{figure}

To determine the cross-sectional area of coronal foot-point in AIA 171 \AA\ image, we obtained the intensity across the coronal foot-point and fitted with a Gaussian with linear function over that intensity distribution as shown in Fig.~\ref{fig:coronal_diameter}. The Full-Width Half Maximum (FWHM) of fitted Gaussian function provides the diameter of the loop at that location in the corona \citep[e.g.,][]{2019A&A...627A..62G}. The obtained diameter is $\approx 1.76\pm0.18\arcsec$ ($1.27\pm0.13$ Mm) for loop 6.  
 
\section{Correlation image in corona}
\label{append:corr171}
\begin{figure}
    \centering
    \includegraphics[width=0.5\textwidth]{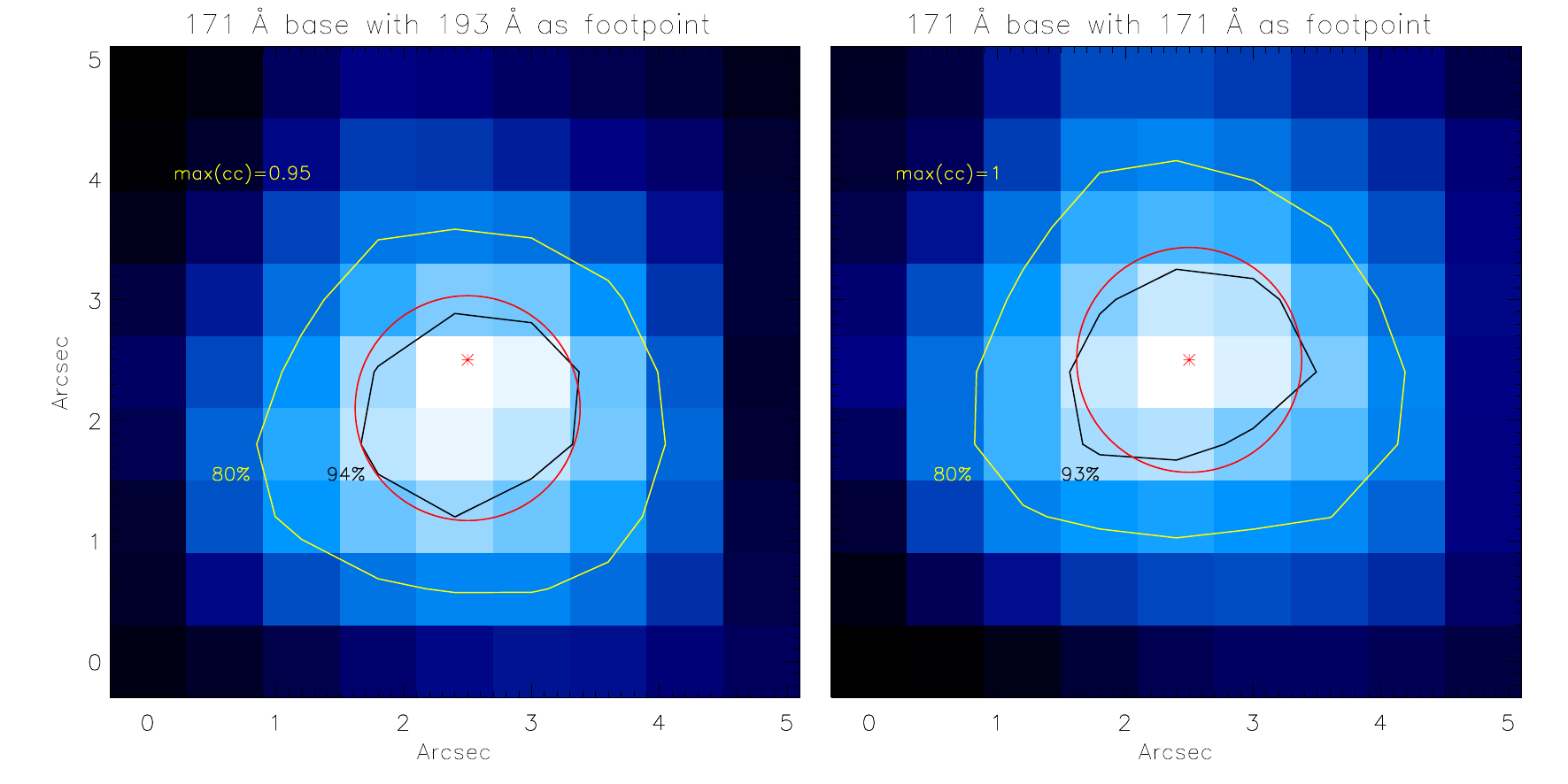}
    \caption{Correlation images obtained in corona as labelled. In each panel, the asterisk sign (*) in the center refers to the coronal foot-point of loop 6 as obtained from AIA 171 \AA\ image. The red circle represents cross-section of the loop obtained from FWHM as described in Appendix~\ref{append:area}. Overplotted yellow colour contours are obtained at 80\% of maximum correlation value. Overlotted black contours as labelled are obtained from maximum correlation value to fit the red circle.} 
    \label{fig:corr171}
\end{figure}

Correlation images obtained between AIA 171 \AA\ image and filtered light curves obtained at coronal loop foot-points in AIA 193 \AA\ (left panel) and 171 \AA\ (right panel) are presented in Fig.~\ref{fig:corr171}.  For details on the method, see Section~\ref{sec:correlation}.

\section{Artefact due to solar rotation}
\label{append:artefact}

In Fig.~\ref{append:artefact}, we show average Fourier power spectra of sunspot umbra obtained by  averaging  individual Fourier powers at each pixel within the umbra. The Fourier power follows a power-law with a strong peak at 3.9 min. This peak is an artefact due to pixel crossing time which happens due to differential rotation of sunspots as described in \citet{2021RSPTA.37900175N}. We calculated the theoretical value of this artefact for our dataset. Analysed sunspot is located at $\approx 7^{\circ}$ south latitude. Using the sswidl routine $diff\_rot$ \citep[based on][]{1990SoPh..130..295H}, we determined the differential rotation rate of this sunspot as 1.88 km~s$^{-1}$. Since 1-pixel (0.6$\arcsec$) here represents 435 km, which leads to the pixel crossing time to be around 3.86 min. This value is same as the peak found in the Fourier power spectra. Also, there will be a slight shift in the latitude for each loop foot-point or locations within the umbra which can shift this artefact by a few seconds. Therefore, we are taking 3.74-4.2 min period window as the spread of this artefact. This artefact is more noticeable when two nearby pixels have significant intensity gradients. 

\begin{figure}
    \centering
    \includegraphics[width=0.45\textwidth]{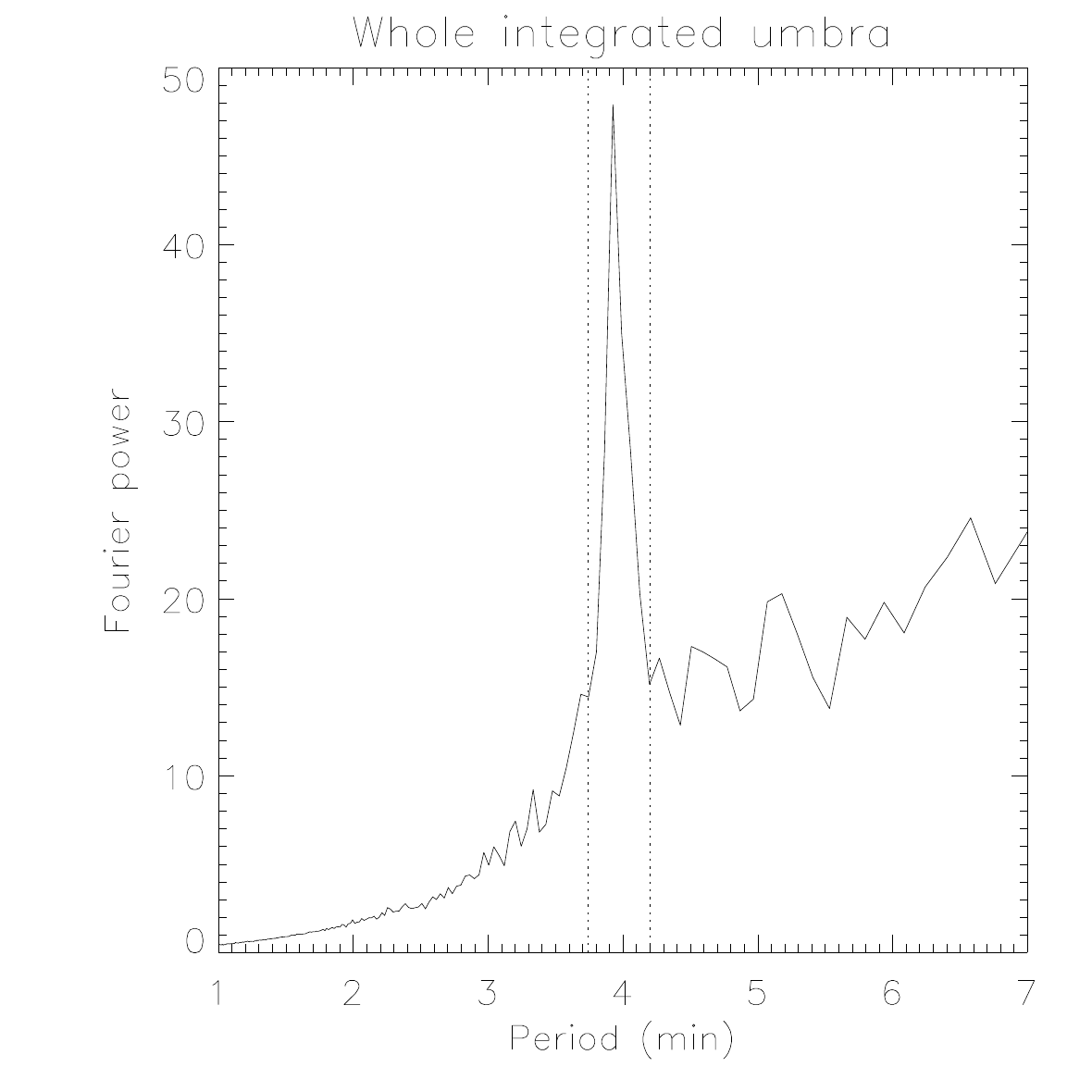}
    \caption{Average Fourier power spectrum of full umbra. The peak at 3.9 min is an artefact corresponding to the ‘pixel-crossing’ time. Two vertical dashed lines at 3.74 and 4.2 min show the period window excluded from our analysis.}
    \label{fig:artefact}
\end{figure}

\section{Temperature profile for umbra}
\label{append:temp}

\begin{figure}
    \centering
    \includegraphics[width=0.49\textwidth]{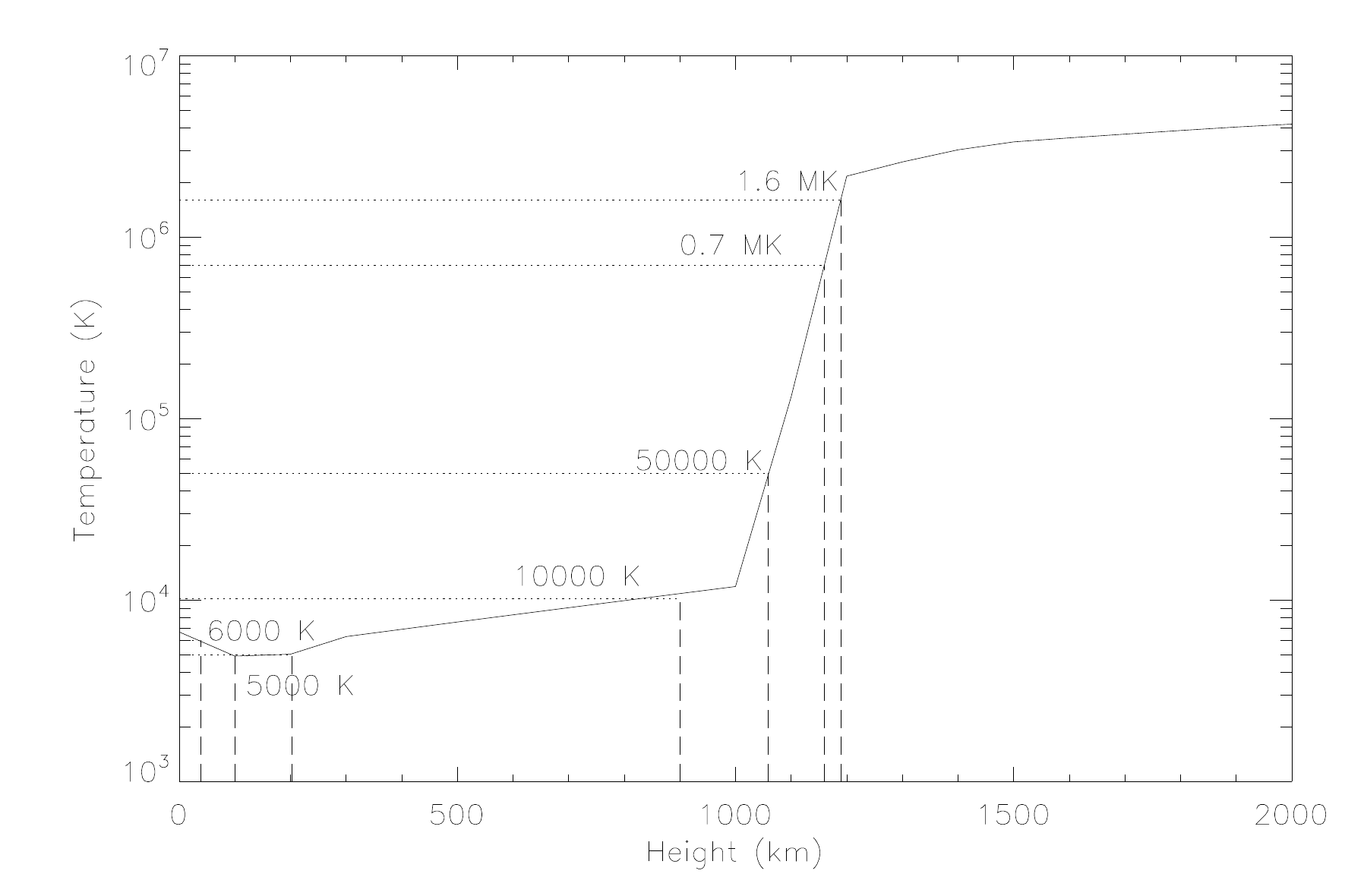}
    \caption{Height-temperature profile for sunspot umbra extracted from \citet{2022FrASS...9.1118D}. Formation heights (denoted by vertical dashed lines) for passbands sensitive to a particular temperature (denoted by horizontal dotted lines) are also provided.}
    \label{fig:height}
\end{figure}

Formation heights in the sunspot umbra for a particular passband sensitive to one temperature is shown in Fig.~\ref{fig:height}. For the purpose, we have utilised umbral model of \citet{2022FrASS...9.1118D} which they claimed close to the atmospheric models of \citet{1999ApJ...518..480F,2009ApJ...707..482F} and others. Here, atmospheric model of \citet{2022FrASS...9.1118D} is preferred over models of \citet{1986ApJ...306..284M} and \citet{2015ApJ...811...87A} due to the higher upper boundary of \citet{2022FrASS...9.1118D}, which includes an extended corona.

\section{Wave Propagation speed for all the loops}
\label{append:wavespeed}

\begin{table*}
    \centering
        \caption{Correlation values and corresponding time lags obtained from Fourier-filtered light curve pairs for all the loops. Wave propagation speeds at different heights are also provided.}
  \label{tab:append1}
  \resizebox{\textwidth}{!}{
    \begin{tabular}{|c|c|c|c|c|c|c|c|c|c|c|c|c|c|c|c|} 
    \hline
\multicolumn{1}{|c|}{Filters (\AA)}   &  
\multicolumn{3}{| c |}{Loop 1 }  &
\multicolumn{3}{|c|}{Loop 2} &
\multicolumn{3}{|c|}{Loop 3} &
\multicolumn{3}{|c|}{Loop 4} & \\
& Correl. coeff. & Time lag  (s) & Obs. speed (km s$^{-1}$) & Correl. coeff. & Time lag  (s)& Speed (km s$^{-1}$)  & Correl. coeff. & Time lag  (s)& Speed (km s$^{-1}$)  & Correl. coeff. & Time lag  (s)& Speed (km s$^{-1}$) \\
\hline
193-171     & 0.53 & -12.0$\pm 12.0$  & --             & 0.96 & 0 $\pm 12.0$     & --             & 0.94 & 0 $\pm 12.0$    & --                  & 0.87 & 0 $\pm 12.0$     & --     \\ \hline
171-304     & 0.60 & 0 $0\pm 12.0$    &--              & 0.93 & 12.0 $\pm 12.0$  & 8.5 $\pm10.7$  & 0.95 & 12.0 $\pm 12.0$ & 8.5 $\pm 10.7 $ & 0.70 & 24.0 $\pm 12.0$  & 4.2 $\pm 3.9 $    \\ \hline
304-2796    & 0.82 & 32.4 $\pm 6.9$   & 4.9 $\pm 2.9$  & 0.86 & 59.9 $\pm 6.9$  & 2.6 $\pm1.2 $ & 0.82 & 53.0 $\pm 6.9$  & 3.0 $\pm 1.37$     & 0.70 & 32.4 $\pm 6.9$   & 4.9 $\pm2.9 $ \\ \hline
2796-1700   & 0.65 & -15.8 $\pm 6.9$  &44.0 $\pm 19.4$ & 0.70& -15.8 $\pm 6.9$  & 44.0 $\pm19.4$& 0.69 & -8.9 $\pm 6.9$  & 78.0 $\pm 60.7$    & 0.77 & -8.9 $\pm 6.9$   & 78.0 $\pm60.7 $  \\ \hline
1700-cont.  & 0.34 & 48.0 $\pm 12.0$ &  3.5$\pm0.9 $   & 0.41 & 24.0 $\pm 12.0$  & 6.96 $\pm 3.51$ & 0.43 & 12.0 $\pm 12.0$ & 13.92 $\pm 13.92$  & 0.24 & 24.0 $\pm 12.0$ &  6.96$\pm 3.51 $\\ \hline
Dopp.-cont. & 0.37 & -48.0 $\pm 12.0$ & 1.29 $\pm 0.33$& 0.61 & -72.0 $\pm 12.0$ & 0.86 $\pm 0.16 $& 0.47 & 2352 $\pm 12$   & --                 & 0.75 & -48.0 $\pm 12.0$ & 1.29 $\pm 0.33 $ \\ \hline
 \end{tabular}}
\end{table*}

\begin{table*}
    \centering
        \caption{Same as Table~\ref{tab:append1}.}
  \label{tab:append2}
  \resizebox{\textwidth}{!}{
    \begin{tabular}{|c|c|c|c|c|c|c|c|c|c|c|c|c|c|c|c|} 
    \hline
\multicolumn{1}{|c|}{Filters (\AA)}   &   
\multicolumn{3}{|c|}{Loop 5} &
\multicolumn{3}{|c|}{Loop 6} &
\multicolumn{3}{|c|}{Loop 7} &
\multicolumn{3}{|c|}{Loop 8} & \\ 
& Correl. coeff. & Time lag $\pm 12$ (s)& Obs. speed (km s$^{-1}$)  & Correl. coeff. & Time lag $\pm 12$ (s)& Speed (km s$^{-1}$)  & Correl. coeff. & Time lag $\pm 12$ (s)& Speed (km s$^{-1}$)   & Correl. coeff. & Time lag $\pm 12$ (s)& Speed (km s$^{-1}$)  \\
\hline
193-171     & 0.95 & -12.0 $\pm 12.0$  & --             & 0.95  & -12.0 $\pm 12.0$   & --                & 0.97  & 0 $\pm 12.0$      & --              & 0.94  & 0 $\pm 12.0$ & -- \\ \hline
171-304     & 0.87 & 12.0 $\pm 12.0$   &8.5 $\pm10.7$  & 0.80  & 24.0 $\pm 12.0$    & 4.2 $\pm3.9 $     & 0.86  & 12.0 $\pm 12.0$   & 8.5 $\pm$10.7  & 0.92  & 12.0 $\pm 12.0$ & 8.5 $\pm10.7 $ \\ \hline
304-2796    & 0.82 & 46.2 $\pm 6.9$    & 3.4$\pm1.6 $ & 0.91  & 46.2 $\pm 6.9$     & 3.4 $\pm1.6 $     & 0.90  & 53.0 $\pm 6.9$    & 3.0 $\pm $1.4  &  0.90 & 46.2 $\pm 6.9$ & 3.4$\pm 1.4$  \\ \hline
2796-1700   & 0.78 & -15.8 $\pm 6.9$   & 44.0$\pm19.4$ & 0.69  & -15.8 $\pm 6.9$    & 44.0 $\pm19.4 $   & 0.81  & -15.8 $\pm 6.9$  & 44.0 $\pm19.4$ & 0.87  & -15.8 $\pm 6.9$ & 44.0$\pm19.4 $ \\ \hline
1700-cont.  & 0.47 & 60.0 $\pm 12.0$   & 2.8$\pm0.5 $  & 0.37  & 48.0 $\pm 12.0$    & 3.5$\pm0.9 $      & 0.49  &  $60.0\pm 12.0$   & 2.8$\pm0.5 $   & 0.46  &  48.0 $\pm 12.0$ &  3.5$\pm$0.9  \\ \hline 
Dopp.-cont. & 0.74 & -48.0 $\pm 12.0$  & 1.3 $\pm0.3 $  & 0.90  & -48.0 $\pm 12.0$   & 1.3 $\pm0.3$      & 0.80  &  $-48.0\pm 12.0$  & 1.3$\pm0.3 $   & 0.80  & -48.0 $\pm 12.0$ & 1.3 $\pm0.3 $ \\ \hline
 \end{tabular}}
\end{table*}

We estimated wave propagation speed at different heights for all the loops as obtained for loop 6. These wave propagation speeds are estimated using time lags obtained from correlation analysis performed on Fourier-filtered light curve pairs and corresponding height difference from the umbral atmospheric model of \citet{2022FrASS...9.1118D} and references therein, see details in Appendix~\ref{append:temp}. In the Tables~\ref{tab:append1} and \ref{tab:append2}, we provide correlation values and corresponding time lags with propagation speeds at that atmospheric height. 

\section{Randomization boot-strap analysis}
\label{append:random}
To check the reliability of correlation coefficient values,  we are performing randomization boot-strap analysis. In this method, it is assumed that in non-periodic time series data, the order in which data is taken is not important. For example, suppose there is no periodicity in the time series. In that case, the order of intensity (or any variable) with respect to time can be changed from $x_1$, $x_2$,... $x_n$ to a randomly arranged $x_{r(1)}$, $x_{r(2)}$,... $x_{r(n)}$, where $n$ is the number of data points and $r(1)$, $r(2)$,... $r(n)$ is a random permutation of the data points \citep[e.g.,][]{2001A&A...368.1095O,2017ApJ...850..206S}. Here we are randomizing the amplitude  modulation curves obtained earlier from different heights. 

In Fig.~\ref{fig:co_am}, pink colour lines are obtained by correlating randomized amplitude modulation curves obtained from the wavelet spectrum of upper height with the randomized amplitude modulation curves obtained from the lower height as labelled using the method described above. Similarly, purple lines are obtained from amplitude modulation curves resulting from the extrema method. We have performed this randomization test 300 times using all the 1200 data points of the time sequence. All the 300 randomized correlation curves are plotted with the same colour lines. All these lines together provide a band for correlation values, and any value beyond this band can be considered as a reliable correlation coefficient value. Here we can see that the band ranges approximately between -0.1 to 0.1, which means that if the correlation coefficient value between the two curves is above 0.1 or below -0.1 then the correlation values can be considered reliable. 

\bsp	

\label{lastpage}

\end{document}